\documentclass[a4paper]{article}
\usepackage[utf8]{inputenc}
\usepackage{jcappub,latexsym,rotating}
\usepackage[normalem]{ulem}
\DeclareMathOperator{\dif}{d\!}

\DeclareMathOperator{\e}{e\!}
\def\mchi{m_{\rm DM}}
\def\sv{\langle\sigma v\rangle}
\DeclareMathOperator{\mx}{X}
\def\mq{\mathcal Q}
\def\mr{\mathcal R}
\def\ms{\mathcal S}

\def\mf{\mathcal F}
\def\mn{\mathcal N}
\def\mh{\mathcal H}
\def\old{{\it old MoI}}
\def\bvec#1{\textrm{\boldmath $#1 $}}

\def\mnras{\rm{MNRAS}}             
\def\sovast{\rm{Soviet Ast.}}
\def\apjs{\rm{Astrophysical Journal, Supplement}}

\title{Universal profiles for radio searches of Dark Matter in dwarf galaxies}
\author[]{Martin Vollmann}\emailAdd{martin.vollmann@tum.de}
\affiliation{Physik Department T31,\\
James-Franck-Stra\ss{}e 1,Technische Universit\"at M\"unchen, \\
D-85748 Garching, Germany}

\abstract{
The phenomenology of diffuse radio emission from Dark Matter annihilation or decay in dwarf spheroidal galaxies is examined.
We introduce (in the context of cosmic-ray physics) a novel strategy for the computation of the relevant synchrotron signals.
In particular, we identify various regimes where, in analogy to prompt gamma rays, the diffuse radio signal from dark matter annihilation/decay can be expressed as the multiplication of a halo times a spectral function. 
These functions are computed here for the first time for a number of benchmark cases.
Furthermore, we find parameter regions in which the emissivity can be well approximated by a universal function $\sim\sin(\pi r/r_h)/r$, where $r$ is the galacto-centric distance and $r_h$ the diffusion-zone radius of the galaxy. 
Our theoretical setup differs from previous work in that, instead of employing a method-of-images strategy, we consider a Fourier-mode expansion of the relevant Green’s functions. 
With this strategy, exact results can be obtained with very low computational cost and for generic dark matter models. 
In particular, $\mathcal O$(10 -- 100) Fourier modes can be easily incorporated into the computations in order to probe the smallest scales of the problem.
We also propose a new strategy to search for dark matter using radio observations of dwarf galaxies that is (1) easy to implement and (2) free of the otherwise large degeneracies in the description of synchrotron signals from dark matter.
Finally, we correct a mistake in a widely used Green's function formula in this context. We show that the original expression leads to systematically incorrect -- and in some cases divergent -- results in  the regime where the characteristic time-scale for diffusion is smaller than that for energy losses.
}

\keywords{dark matter, dwarf galaxies, synchrotron radiation}
\begin{document}

\begin{flushright}
{\small
TUM-HEP-1298/20\\
April 27th, 2020
}
\end{flushright}

\maketitle

\section{Introduction}
The identification of the nature of the dark matter is one of the most fundamental goals in the current scientific context.
In particular, the study of possible non-gravitational interactions of dark matter has become a very active field in the last several decades.
Within the $\Lambda$CDM framework it is possible to explain the observed DM abundance in several ways. 
However, the most investigated paradigm to this date is undoubtedly that where the DM is made of what is known as weakly-interacting massive particles (WIMPs) \cite{Jungman:1995df}. 
These hypothetical particles are such that their number density freezes out when the expansion rate of the Universe becomes larger than their annihilation rate. 
Intense efforts in searching for these particles in the present is motivated by a very favorable experimental situation. 
Indeed, current and next-generation experiments have reached and will reach high enough sensitivities to (directly or indirectly) detect several benchmark WIMP realizations. 
For a review on the current situation and future prospects, we refer to e.~g. \cite{PerezdelosHeros:2020qyt,Leane:2018kjk,Bertone:2018xtm,Arcadi:2017kky,Kahlhoefer:2017dnp,Bramante:2015una}.
In particular, dedicated observations of dwarf spheroidal satellite galaxies (dSph) of the Milky Way (MW) using gamma rays by Fermi-LAT \cite{Ackermann:2015zua,Ahnen:2016qkx} yield the strongest bounds on the annihilation cross sections for a few to hundred GeV scale WIMP models.
The interest in these satellite galaxies is motivated by their prevalent DM content and low backgrounds.

Nevertheless, as the presently running experiments have only been able to exclude several benchmark scenarios for the WIMP, complementary approaches are necessary. 
One natural possibility is, of course, paying closer attention to alternative (non-WIMP) hypotheses (axions, sterile neutrinos, self-interacting DM, etc.) which is reflected in the increase on theoretical and experimental work on WIMP alternatives in recent years. The second strategy, which we pursue in this article, is digging further into unexplored avenues in the search for WIMPs. 
In particular, the main purpose of this paper is to provide a novel picture on how theoretical predictions of radio fluxes from dark matter annihilation in dwarf galaxies are performed, with the goal of searching for such signals using existing and future radioastronomical observations. 
Moreover, our results can be trivially extended to non-WIMP scenarios where the DM can be arbitrarily heavy but unstable. 
In this case, it is the decay of the DM particles in the dSphs (rather than their annihilation) the process of interest.

The radio signals in which we are interested are generated by the motion of very energetic electrons and positrons that were produced by the annihilation/decay of DM particles in the presence of magnetic fields.
Due to our ignorance about the magnetic field and its turbulence in dwarf galaxies, the theoretical description for these {\it synchrotron radiation} signals can be uncertain by several orders of magnitude. 
This is in strong contrast with the relevant phenomenology for e.~g. the prompt-emission searches in gamma-ray frequencies --the golden channel for indirect detection of DM-- where the uncertain magnetic fields do not play any role.

Early work \cite{Berezinsky:1992mx,Berezinsky:1994wva,Gondolo:2000pn,Bertone:2001jv,Aloisio:2004hy} (also own work \cite{Asano:2012zv,Bringmann:2014lpa}) focused on setting up the theoretical grounds for putting bounds on neutralino DM models using the Galactic Center as target, without conducting any tailored experiments but rather comparing with existing data.
Similarly, in \cite{Tyler:2002ux,Colafrancesco:2006he} the synchrotron emission by DM annihilations in the Draco dwarf was modelled for the first time and compared to Very Large Array (VLA) data that was available at the time.
Fortunately, this situation has been changing in recent years.
Namely, in \cite{Spekkens:2013ik,Natarajan:2013dsa,Regis:2014joa,Regis:2014koa,Regis:2014tga,Natarajan:2015hma, Regis:2017oet, Kar:2019hnj, 2020MNRAS.494..135C,2020MNRAS.496.2663V,Bhattacharjee:2020phk} observational efforts have been undertaken to search for DM annihilation signatures in dSphs. 
The upcoming Square Kilometre Array consortium includes the search for these synchrotron signals among their scientific interests \cite{2013arXiv1301.4124A} and several forecasts of its capabilities have been carried out, e.~g. \cite{Kar:2018rlm,Beck:2019ukt,Kar:2019cqo,Cembranos:2019noa}.
As further motivation, the lack of positive signals of DM by the more ``conventional'' searching strategies demands an intensification of the efforts to look for it by means of already available infrastructures, such as the several existing radio telescopes. 
Moreover, the DM search program using radio observations offers valuable information on the nature of the DM which is complementary to what we have already learned about it by considering other messengers such as gamma rays, cosmic-ray antiparticles, neutrinos, etc. 

In this work we aim at improving the methodology that are relevant for predicting synchrotron signals from DM annihilation/decay in the satellite dSphs.
Our attention is put on the theoretical modelling of the relevant comic-ray transport in these galaxies.
In particular, we identify three distinctive regimes (A, B and C) which are defined in terms of the two characteristic time scales in the problem: diffusion and energy-loss time scales.
Approximate (effective) descriptions that are appropriate in each regime are introduced and put under scrutiny.
A common property shared by these approximations is the fact that the emissivity from DM annihilation or decay in these approximations can be expressed as the product of a halo times a spectral function.

We provide these functions for generic DM particle models and density profiles and discuss the transition between the characteristic regimes of the problem.
Our computations are done in such a way that the interested reader/user will be able to (1) obtain the theory input that is relevant for their analyses in a computationally efficient way and (2) gain a deeper intuition on all the physical aspects of the signal. 
Furthermore, due to the large degeneracies that we are able to identify in the problem, we propose a novel DM-search strategy that minimizes all associated redundancies in e.~g. model-parameter scans.
In particular, in Ref. \cite{2020MNRAS.496.2663V} some of these methods were used in putting constraints on the DM annihilation cross section using observations of the Canes Venatici I dwarf spheroidal (dSph) galaxy.
Our methods are transferable to modelling the inverse Compton scattering X-ray (soft gamma-ray) signal in dSphs with direct impact on indirect detection of WIMPs, decaying DM and even primordial black holes (see e.~g. \cite{Dutta:2020lqc}).
In return for the comprehensive attention we paid on all details, we also partly correct results that have been employed in previous works.
Namely, we detected a mistake in the formula for the Green's function adopted in Refs. \cite{Colafrancesco:2005ji, Colafrancesco:2006he, Jeltema:2008ax, Yuan:2009yy, Calvez:2010jq, Huang:2010yga, Spekkens:2013ik, Natarajan:2013dsa, Colafrancesco:2014coa, Natarajan:2015hma, Beck:2015rna, Marchegiani:2016xyv, McDaniel:2017ppt, McDaniel:2018vam, Kar:2018rlm, Beck:2019ukt, McDaniel:2019niq,Kar:2019cqo, Cembranos:2019noa, Kar:2019hnj, Bhattacharjee:2019jce, Kar:2020coz,Dutta:2020lqc} in order to solve the relevant transport equation for cosmic-ray electrons.

The article is organized as follows. 
In section \ref{sec:theory} we discuss the basic theory of synchrotron radiation and the cosmic-ray transport model that we adopt.
We then move on introducing the Green's-function solution for the cosmic-ray electron density and the three approximations that we are able to develop in the three regimes A, B and C which we also introduce there.
In section \ref{sec:universal} we show our simplified formulas for the emissivity and the definitions of our regime-dependent halo and spectral factors.
This is followed by section \ref{sec:numerics} where we first describe the most characteristic features of the halo functions and then discuss our numerical strategies in computing the spectral functions for generic DM models.
The section is concluded with a series of tests that we conducted in order to demonstrate the validity and usefulness of our approximations and further discussions of our results.
In section \ref{sec:draco} we explain with an example the proposed strategy to search for DM using radio observations of dwarf galaxies.
We then conclude. 
Supplementary material such as some derivations and analytic formulas for the brightness halo functions and some definitions can be gleaned from appendices \ref{app:analytics} and \ref{app:brighthalos}.
Appendix \ref{app:spectra} contains the spectra for a selection of DM models.
In appendix \ref{app:bench} we perform further checks on the performance of our approximations and in \ref{app:moi} we verify that our results reproduce those that are obtained by employing the method of images.
We also assess there the numerical impact the mistake made in previous works on that formula for the Green's function.
Finally, we provide a detailed derivation of \eqref{eq:bounds} in appendix \ref{app:bounds}. 
Eq. \eqref{eq:bounds} is namely the equation that we use in order to estimate the bounds that can be set on e.~g. the annihilation rate of WIMP DM by a given radio observation of dwarf galaxies.

\section{Synchrotron signals from dwarf galaxies}
\label{sec:theory}
The theory of synchrotron radiation in the context of astrophysics is firmly established and we refer non-expert readers that are interested in the details to textbooks on this matter (e.~g. \cite{1976PhuZ....7S.128.,1979rpa..book.....R,1992hea..book.....L} and \cite{MartinVollmann:2015gea} in the context of DM). 
For our purposes we assume that the magnitude of the turbulent magnetic field is constant and randomly oriented all over the galaxy and that the absorption of radio waves in the dwarf galaxy is negligible. 
In this case, the synchrotron brightness $I_\nu$ is given by
\begin{equation}
\label{eq:simpbright}
I_{\nu} =\int_0^\infty\frac{\dif z}z\mf(z)\,I_{\frac{\nu}z}^{\rm mn}\ ,
\end{equation}
where \cite{1973plas.book.....K} 
\begin{eqnarray}
\label{eq:fav}
\mf(z)&=&z^2\left(K_{\frac43}(z)K_{\frac13}(z)-\frac35\left[K_{\frac43}^2(z)-K_{\frac13}^2(z)\right]\right)\\
I_{\nu}^{\rm mn} &=& \int_{\rm l.o.s.}\!\dif l\, j_{\nu}^{\rm mn}(\bvec r(l))\ ,
\end{eqnarray}
``l.~o.~s.'' stands for {\it line of sight}, $K_{4/3}(z)$ and $K_{1/3}(z)$ are modified Bessel functions of the second kind (also Macdonald functions) and $j_\nu^{\rm mn}(\bvec r)$ is the ``monochromatic'' synchrotron emissivity. 
This is given by
\begin{equation}
j_{\nu}^{\rm mn}(\bvec r)=\frac{\sqrt3 e^3 B}{m_e}E(\nu)\left[\frac1{4\pi}n_{e^-}(\bvec r,E(\nu))+\frac1{4\pi}n_{e^+}(\bvec r,E(\nu))\right]\ ,
\label{eq:mnemiss}
\end{equation} 
where $n_{e^-}$ ($n_{e^+}$) is the number of electrons (positrons) per unit energy and volume. 
In the language of the monochromatic approximation, $E(\nu)=\sqrt{2\pi m_e^3\nu/(3eB)}$ is the required energy for a synchrotron emitter (electron or positron) to radiate at frequency $\nu$.
This picture is of course very useful for gaining insights into the problem but it can be misleading.
In the most rigorous sense the so-called ``monochromatic approximation'' is not an approximation per se. 
More specifically, the synchrotron-power spectrum can not be deformed into Dirac deltas by taking any limits on the relevant parameters. 
In spite of this, the monochromatic ``approximation'' is routinely used in the literature in order to obtain quick order-of-magnitude estimates.

Regarding unit conventions, we use $c=k_B=1$ so that the brightness temperature $T_B$ and the brightness $I_\nu$ are related by $I_\nu=2\,\nu^2\,T_B$.
The flux density, on the other hand, reads
\begin{equation}
\label{eq:fluxdsimp}
S_\nu=\int\dif\Omega\cos\theta I_\nu\simeq\cos\theta_0 \int\dif\Omega\int\dif l\, j_\nu\simeq\frac{\cos\theta_0}{R^2}\int\dif V\,j_\nu\ ,
\end{equation} 
where $R$ is the distance to the object and $j_\nu=\int\frac{\dif{z}}z\mf(z)j_{\nu/z}^{\rm mn}$.
The angle $\theta_0$ is the angle with respect to the zenith of the surface that is being radiated upon.
We will assume henceforth that $\theta_0=0$ and that the volume integral \eqref{eq:fluxdsimp} is performed all over the diffusion zone of the galaxy, i.~e. the region of the galaxy where the turbulent nature of the magnetic field leads to a diffusive propagation of the cosmic-ray electrons and positrons.

Having established our conventions, we can now turn our attention to the determination of the cosmic-ray electron number density. 
This is the central ingredient in the computation of the emissivity \eqref{eq:mnemiss} and a very challenging task. 
It requires (1) the modelling of the rather uncertain \emph{transport} of cosmic-ray electrons and positrons (henceforth CREs) in the dSphs, (2) a \emph{mass model} for the galaxy and (3) understanding the injection of electrons per \emph{annihilation/decay} within a given WIMP model. 
In this work we focus on point (1) while we refer to the vast existing literature on points (2) and (3).
Concretely, concerning point (2), the distribution of DM in dSphs can be inferred from the stellar-kinematics data that is available for each galaxy.
Specifically, given a parametrization for the DM density $\rho(r)$ in the dSph, e.~g. Einasto\cite{1965TrAlm...5...87E}, (generalized) \cite{Hernquist:1990be,Dehnen:1993uh,Zhao:1995cp} Navarro-Frenk-White (NFW) \cite{Navarro:1996gj}, Burkert\cite{Burkert:1995yz}, etc.\footnote{These parametrizations are motivated by cosmological ($\Lambda$-Cold-DM) N-body simulations.} a Jeans-based likelihood method is adopted in order to fit the relevant free parameters of the aformentioned parametrizations. 
In particular, in the Fermi-LAT search for DM using dSphs \cite{Ackermann:2015zua}, the results from \cite{2015MNRAS.451.2524M} were used. 
However, recent work in this direction is rather extensive. 
See e.~g. \cite{Chang:2020rem,Horigome:2020kyj,Alvarez:2020cmw,Chiappo:2018mlt,Pace:2018tin} and references therein.

Regarding point (3), it is standard practice to parametrize the production by DM annihilation or decay of stable particles such as electrons, protons, gamma rays, etc., in terms of a set of the model-specific branching ratios into two-body standard-model (SM) states: $b\bar b$, $W^+W^-$, $\tau^+\tau^-$, etc. and their associated stable-particle yields.
The latter can be obtained by dedicated Monte-Carlo event generators such as {\sc Pythia}\cite{Sjostrand:2014zea}, {\sc Herwig}\cite{Bellm:2015jjp} or directly connected with indirect detection of DM: DarkSUSY\cite{Bringmann:2018lay}, MicrOMEGAs\cite{Belanger:2018ccd}, PPPC\cite{Cirelli:2010xx}.  
This approach is very practical and has a rather broad applicability on generic models of particle DM. 

We now direct all our attention to the modelling of the propagation of the CREs in the dSph. 
In virtue of the large uncertainties present in the problem, a simplified picture in describing the CRE transport becomes appropriate.
Inspired by the well-tested diffusion-loss transport model for the CREs in our MW Galaxy\cite{1976RvMP...48..161G}, a CRE propagation model that is adapted to the approximate spherical symmetry of the dwarf galaxies has been considered in \cite{Colafrancesco:2005ji, Colafrancesco:2006he, Jeltema:2008ax, Yuan:2009yy, Calvez:2010jq, Huang:2010yga, Spekkens:2013ik, Natarajan:2013dsa, Colafrancesco:2014coa, Natarajan:2015hma, Beck:2015rna, Marchegiani:2016xyv, Leite:2016lsv, McDaniel:2017ppt, McDaniel:2018vam, Kar:2018rlm, Beck:2019ukt, McDaniel:2019niq,Kar:2019cqo, Cembranos:2019noa, Kar:2019hnj, Bhattacharjee:2019jce,2020MNRAS.496.2663V, Kar:2020coz}.
In this approach the use of semi-analytical methods is possible and we will explore these in the remaining sections of the paper.

\subsection{Transport model}
As argued above, we will adopt a very simple transport model that is described by the following diffusion-loss equation   
\begin{equation}
\label{eq:diffloss}
D(E)\frac1{r}\frac{\partial^2}{\partial\, r^2}(r\,n_e)+\frac{\partial\, }{\partial E}[b(E)n_e]+s(r,E)=0 \quad,\qquad n_e(r_h,E)=0\ ,
\end{equation}
where $n_e(r,E)$ is the CRE density and $D(E)$, $b(E)$, $s(r,E)$ the diffusion, energy-loss and particle-injection coefficients respectively. 
We thus assume \emph{stationarity}, \emph{spherical symmetry} and \emph{isotropic momentum distribution}: all terms in the equation can only depend on the galactocentric distance, $r$, and the energy, $E$, of the CREs.
See \cite{2019MNRAS.488.1401M} for a discussion on reacceleration effects in clusters of galaxies and why these are irrelevant in dSphs.
The validity of the stationarity is further discussed below in \ref{sec:eloss}. 
A stronger assumption is the fact that the energy-losses and diffusion coefficients do \emph{not} depend on the galactocentric distance $r$.
This assumption which is also adopted in conventional CR transport models in the MW \cite{1976RvMP...48..161G}, is closely related to the constancy of the turbulent magnetic field within the diffusive volume of the galaxy. 
This condition is also necessary for the validity of \eqref{eq:simpbright}.
The somewhat more general case where diffusion, energy-loss coefficients and magnetic fields can be made space dependent (within the spherical-symmetry assumption) is considered in \cite{Regis:2014koa,Regis:2014tga,Regis:2017oet} and discussed in the appendix A of \cite{Regis:2014koa}. 
A numerical approach: the Crank-Nicholson algorithm\footnote{This method has been implemented in widespread Galactic CR propagation codes GALPROP \cite{Strong:1998pw} and DRAGON \cite{Evoli:2016xgn,Evoli:2008dv}.}, is necessary in that case.

In analogy with the CRE propagation modelling in the MW, the boundary condition $n_e(r_h,E)=0$ introduces a length-scale parameter in the problem: the diffusion radius of the halo $r_h$. 
This length scale is physically understood as the radius above which the CRE can freely stream. 

\subsection{Injection}
For annihilating DM, the rate per unit energy $s(r,E)$, at which CR electrons \emph{or} positrons with energy $E$ are produced at a given point with a galactocentric distance $r$, is given by
\begin{equation}
\label{eq:inj}
s(r,E)=\left.\frac1{2\mchi^2}\rho^2(r)\frac{\dif\,\sv}{\dif E}\right|_{\chi\chi\to e^-+X}\ ,
\end{equation}
where we implicitly assumed that the DM particle is its own antiparticle. 
Otherwise, the formula has to be divided by an additional factor of 2. 
In this formula, $\mchi$ is of course the mass of the WIMPs, $\rho(r)$ the DM-density function of the dSph. 
The last term is the velocity-averaged differential \emph{annihilation rate} for semi-exclusive annihilation of the DM pair into electrons with energy $E$ and all possible $X$ ``sister'' particles. 

In the case of decaying DM, the formula reads
\begin{equation}
\label{eq:decinj}
s(r,E)=\left.\frac1{\mchi}\rho(r)\frac{\dif\,\Gamma}{\dif E}\right|_{\chi\to e^-+X}\ ,
\end{equation}
where $\Gamma$ is the (velocity-averaged) semi-exclusive decay rate of the DM into electrons and the rest of the terms ($\mchi$, $\rho(r)$, $E$ etc.) are analogous to the former case.

For future purposes, let us define spatial and spectral injection functions ($\mr(r)$ and $\ms(E)$ respectively) such that\footnote{For WIMPs, the factorization of $s(r,E)$ as the product of a function that only depends on $r$ and another function that only depends on $E$ assumes the standard $s$-wave (or pure $p$-wave) annihilation.} $s(r,E)=\mq\,\mr(r)\,\ms(E)$.
Concretely,  $\mr(r)=\rho^2(r)$ ($\rho(r)$) and $Q=\sv/(2\mchi^2)$ ($\Gamma/\mchi$) for annihilating (decaying) DM.
As already discussed, there are several benchmark models for $\rho(r)$ (and thus $\mr(r)$) that are motivated by cosmological $\Lambda$CDM N-body simulations. 
Since our focus here is the CRE transport, we will consider the following two extreme-case simplified halo-mass distributions:

\subsubsection{Point mass (Dirac-delta) profile}   
For annihilating DM this is defined by
\begin{equation}
\mr_{\rm pm}(r)=\frac{K_{\rm pm}}{4\pi r^2}\delta(r)\ ,
\label{eq:dd}
\end{equation}
where the relationship between the normalization constant $K_{\rm pm}$ and the ``$J$-factor''\footnote{$J=\iint\dif\Omega\dif l\,\rho^2$} of the galaxy is evident: $J_{\rm pm}=K_{\rm pm}/R^2$. 
This ansatz describes the case of an infinitely-concentrated DM halo. 
For decaying DM, the definition is analogous: $K_{\rm pm}$ is substituted by $M_h$, the total mass of the halo. 

\subsubsection{Constant-density  profile}
In this case the DM density function is defined as 
\begin{equation}
\rho_{\rm co}(r)=\rho_h \ , \ r<r_h\ ,
\label{eq:core}
\end{equation}
for both annihilating and decaying DM. 
Thus, $\mr_{\rm co}^{\rm ann.}=\rho_h^2$ and $\mr_{\rm co}^{\rm dec.}=\rho_h$. 
This model of course encompasses all \emph{cored profiles} such that their core radii satisfy $r_{\rm core}>r_h$.
In the next sections it will become clearer why, besides being enormously simplifying, these extreme choices for $\rho(r)$ capture the main phenomenologically interesting aspects of performing halo variations on the synchrotron signal. 
Certainly, the more realistic and thus phenomenological interesting halo parametrizations (see first column of table \ref{tab:halodec}) are also considered in detail in this paper.

\subsubsection{CRE yields from annihilation/decay}
Concerning the microphysics part of \eqref{eq:inj}, we adopt the standard parametrization (e.~g. \cite{Bringmann:2018lay}) which, for annihilating DM, reads:
\begin{equation}
\label{eq:svexp}
\ms(E)=\frac1{\sv}\left.\frac{\dif\,\sv}{\dif E}\right|_{\chi\chi\to e^-+X}=\frac1{\sv}\sum_{ab}\sv_{\chi\chi\to ab}\frac{\dif\,N_{ab}}{\dif E}=\sum_{ab}{\rm BR}_{ab}\frac{\dif\,N_{ab}}{\dif E}\ ,
\end{equation}
BR stands for branching ratio, $ab=\{e^+e^-,\mu^+\mu^-,\tau^+\tau^-,\ldots\}$ and $\langle\sigma v\rangle$ is the total annihilation rate. 
As mentioned above, the electron yields $\dif\,N_{ab}/\dif E$ are typically obtained by Monte-Carlo methods. 
In this framework, the initial center of mass energy of the annihilation or the mass of the decaying DM particle captures all the information about the initial state on which these electron yields can depend.
Therefore, all electron yields from annihilation of DM particles of a given mass $m_0$ correspond to electron yields of decaying DM with a mass of $2m_0$.
Simpler expressions certainly exist if e.~g. the leading-process (tree-level) approximation is adopted. 
For instance, $\frac{\dif\,N^{\rm tree}_{e^-e^+}}{\dif E}=\delta(E-E')$ where $E'=\mchi$ for DM annihilation in the $\mchi\gg m_e$ limit.

As we will argue below, the structure of our transport equation \eqref{eq:diffloss} is such that, once their solutions for the particular case of Born-level $e^+e^-$ electron injection are known \emph{for arbitrary} DM mass, then the general solution for any particle-DM model can be trivially obtained.
We may thus focus on the monochromatic electron injection $\frac{\dif\,N^{\rm tree}_{e^-e^+}}{\dif E}$ without losing generality.

\subsection{Diffusion}
In analogy with the Milky Way, we model the energy dependence of the diffusion coefficient $D(E)$ as a power-law distribution: $D(E)=D_0(E/E_0)^\delta$, where $E_0$ is arbitrary and, as is customary, we adopt $E_0=1$~GeV.
The remaining parameters $\delta$ and $D_0$ are assumed to be constant and independent from each other\footnote{Keep in mind, however, that within a given MHD turbulence model (fixed $\delta$) the normalization parameter $D_0$ can be determined once the ``smooth'' component (turbulent) magnetic field $B$ and its correlation length $L_c$ are known (see e.~g. \cite{Strong_2007}).
This is of course beyond the scope of this article and the reason why we treat $D_0$, $B$ and $\delta$ independently.}.
This parametrization is consistent with several benchmark models of magnetohydrodynamic (MHD) turbulence.
Specifically, $\delta=1/3$, 1/2 for Kolmogorov\cite{1941DoSSR..30..301K} and Kraichnan \cite{Kraichnan:1965zz} turbulence respectively.
 
In order to get a feeling about the possible values the normalization parameter $D_0$ can assume, we can consider two bounds. 
On the small $D_0$ end, we have the Bohm limit which is such that $\delta=1$ and $D(E)=r_L/3=E/(3eB)\simeq3\times10^{22}\,{\rm cm}^2/{\rm s}\times(E/{\rm GeV})(B/\mu{\rm G})^{-1}$. 
On the opposite (free-streaming) limit, $\delta=0$ and $D=\lambda_{\rm mean}/3\simeq 3\times10^{31}\,{\rm cm}^2/{\rm s}\times(\lambda_{\rm mean}/{\rm kpc})$ where $\lambda_{\rm mean}$ is the mean free path of the electron which is formally infinity in the free-streaming limit.
In the MW, however, recent analyses (e.~g. \cite{Weinrich_2020,Wu_2019,Evoli_2019,Johannesson:2016rlh,Korsmeier_2016}) using several CR species prefer values of $\sim\mathcal O(10^{28}\,{\rm cm}^2/\rm s)$ for $D_0$ and $\delta=0.3 - 0.7$, which we adopt in section \ref{sec:numerics} as reference.

Finally, the diffusive transport picture in the dSph is only complete once the diffusion-radius parameter $r_h$ and the turbulent magnetic field $B$ are provided.
The former plays an analogue role as the thickness and radius parameters in the standard cylindrical picture of the MW.
By construction, $r_h$ has to be of the order of the measurable half-light radii $r_\star(\sim\,0.1$~--~1~kpc) of these galaxies.
The introduction of $r_h$ enables us to define a typical time scale of diffusion $\tau_{\rm diff.}=r_h^2/D(E)$, which in the most extreme scenarios can in principle assume values that are as short as 10$^4$~yr and as long as $\sim$~10$^3$~Gyr for $r_h$ of $\mathcal O$(kpc).
In fact, it will prove rather useful for the discussion of our results to introduce the parameter $\tau_0=r_h^2/D_0$.
Concretely, rather than considering independent variations of the $D_0$ and $r_h$ parameters, we will vary $\tau_0$ and $r_h$ simultaneously.

The turbulent magnetic field parameter $B$ in the dSphs, on the other hand, is even more uncertain. 
It could --at least in principle-- be suppressed by several orders of magnitude with respect to the $\mathcal O(1-10\mu$G) magnetic-field strengths that were typical when there was star formation in those galaxies.
However, estimates of the typical Ohmic dissipation time scales for this type of frozen-in field configurations\footnote{The magnetic diffusivity is assumed to be purely turbulent: $\beta\simeq\frac13L_cv\simeq10^{26}{\rm cm}^2/{\rm s}\times(L_c/100{\rm pc})(v_c/{\rm 10~km/s})$, where $v_c$ is the characteristic velocity of turbulent motions.} $\tau_{\Omega}\simeq r_h^2/\beta=0.3\,{\rm Gyr}(r_h/{\rm kpc})^2(\beta/(10^{26}{\rm cm}^2/\rm s))^{-1}$ \cite{1988ASSL..133.....R}, suggest that the $\mathcal O$(few $\mu$G) magnetic-field strengths that were present in these type of galaxies during their star-forming epochs, continue to exist for time periods that are comparable with the current age of a typical dSph.
A similar argument and a much longer discussion on the magnetic properties of present-day dSphs is given in \cite{Regis:2014koa}.

\subsection{Energy losses}
\label{sec:eloss}
The relevant processes for energy loss are synchrotron radiation itself and inverse Compton scattering (ICS) off the CMB photons. 
Interactions with thermal electrons and other matter components are suppressed by their small \cite{Regis:2014koa} ($\mathcal O(10^{-6}$cm${}^{-3})$) number densities in the dwarf galaxies\footnote{Relatedly, the smallness of the electron and ion densities in the dSphs explains why it is appropriate to neglect those absorption effects (free-free and electron scattering) that are relevant for the propagation of radio waves in the dSphs. Namely, typical values of the mean free path of radio waves in such galaxies $\sim1/(n_e\sigma_T)=500$~Gpc~$\times(10^{-6}$cm${}^{-3}/n_e)$ ($\sigma_T$ is the Thomson-scattering cross section) are several orders of magnitude larger than the size of the galaxy.}.
The contribution from these processes to the energy-loss parameter are given by\cite{1979rpa..book.....R}
\begin{eqnarray}
\label{eq:synchloss}
b_{\rm synch}(E) &=&\frac{4\,e^4B^2E^2}{9m_e^4}\\
{}&\simeq&2.5291\times10^{-18}\,{\rm GeV/s}\left(\frac{B}{1\,\mu\rm G}\right)^2\left(\frac{E}{1\,\rm GeV}\right)^2\\
\label{eq:ics}
b_{\rm ICS}(E) &=& \frac{b_{\rm sync}(E)}{u_{\rm mag}}u_{\rm rad}\ ,\ u_{\rm mag}=\frac{B^2}{8\pi}\ ,
\end{eqnarray}
where for the radiation field we assume that only CMB photons contribute: $u_{\rm rad}=4\,\sigma_{\rm S}T_{\rm CMB}^4=0.25$~eV/cm$^3$. 
The effects of further radiation fields other than CMB on the IC scattering can be estimated in terms of the dSphs' V-band luminosities by using $u_\star/u_{\rm CMB}\simeq L_V/(4\pi r_\star^2)/u_{\rm CMB}\approx$~0.3\% $\times(L_V/10^6L_\odot)\times$(1kpc/$r_\star)^2$. Thus, we may assume that CMB photons are responsible of all ICS losses and consequently,
\begin{eqnarray}
b(E) &=&b_{\rm ICS}(E)+b_{\rm sync}(E)=\frac{32\pi e^4 E^2}{9m_e^4}(u_{\rm rad}+u_{\rm mag})\\
{}&\simeq&2.546\times10^{-17}\,{\rm GeV/s}\left[1+\left(\frac{B}{3.173\,\mu\rm G}\right)^2\right]\left(\frac{E}{1\,\rm GeV}\right)^2\ .
\end{eqnarray}

Note that for energies $E\lesssim\mathcal O(0.1$~GeV), the typical energy-loss time scale $\tau_{\rm loss}\sim E/b(E)\sim12.4\,{\rm Gyr}\, (E/{\rm GeV})^{-1}$ is comparable with (larger than) the age of the galaxy. 
Therefore, the validity of the stationarity condition (in the regimes A and B defined below) would then not be valid for such sub-GeV electron energies.
In particular, the effects of e.~g. the galaxy halo formation history could play a role in the prediction of synchrotron fluxes for DM masses below $\mathcal O$(100~MeV).
On the other end (heavy DM), the ICS formula \eqref{eq:ics} is valid in the Thomson regime: $2E_{\rm CMB}E\ll m_e^2$ which is means that the formula is valid up to energies $E\lesssim \mathcal O($100~TeV). 
Since accounting the aforementioned effects goes beyond the goal of this paper we focus on the 1~GeV~$\lesssim\mchi\lesssim$~100~TeV range. 

\subsection{Green's function solution}
Next, we discuss the semi-analytical solution for eq. \eqref{eq:diffloss}. 
This is obtained by means of the Green's function method. 
Specifically, by introducing the Syrovatskii variable \cite{1959SvA.....3...22S} $\lambda^2$, which we define as
\begin{equation}
\lambda^2(E)=\int_E^{\infty}\dif E'\frac{D(E')}{b(E')}\ ,
\label{eq:syrovatskii}
\end{equation}
and some field redefinitions, it is possible to prove that eq. \eqref{eq:diffloss} acquires the form of a \emph{one-dimensional} heat equation.
The variable $\lambda$ has length units and by using our expressions for $D(E)$ and $b(E)$, it is straightforward to arrive at the following formula:
\begin{eqnarray}
\lambda^2(E) &=&\frac{9m_e^4D_0E_0^{-\delta}}{32\pi e^4u_{\rm EM}}\int_E^{\infty}\dif E'E'^{-(2-\delta)}=\frac{9m_e^4D_0}{32\pi e^4u_{\rm EM}(1-\delta) E_0}\left(\frac{E_0}{E}\right)^{1-\delta}\nonumber\\
{}&\simeq&(6.42\,{\rm kpc})^2\left(\frac{D_0}{10^{28}\,\rm cm^2/s}\right)\frac1{1-\delta}\frac1{1+\left(\frac{B}{3.135\,\mu\rm G}\right)^2}\left(\frac{1\,\rm GeV}{E}\right)^{1-\delta}\ .
\label{eq:syrint}
\end{eqnarray}

It will prove useful, however, to introduce the dimensionless variable $\eta(E)\equiv\pi^2\lambda^2(E)/r_h^2$: 
\begin{equation}
\eta(E)\simeq(3.52)^2\left(\frac{\tau_0}{1\,\rm Gyr}\right)^{-1}\frac1{1-\delta}\frac1{1+\left(\frac{B}{3.135\,\mu\rm G}\right)^2}\left(\frac{1\,\rm GeV}{E}\right)^{1-\delta}\ ,
\label{eq:etavariable}
\end{equation}
where the parameter $\tau_0=r_h^2/D_0$ was introduced above.
In particular, $\lambda\propto E^{-1/3}$ ($E^{-1/4}$) in the Kolmogorov (Kraichnan) case, which shows that for our problem, the Syrovatskii variable $\lambda(E)$ is largely insensitive to large changes in $E$.
However, we will see that even small changes in the Syrovatskii variable get amplified by the strong (exponential) $\lambda$-dependence of the solutions for \eqref{eq:diffloss}. 
Specifically, 
\begin{equation}
\label{eq:masterdiffloss}
n_e(r,E)=\frac1{b(E)}\int\dif E'\int\dif r'\frac{r'}{r}G_{r_h}^{\rm 1D}(\Delta\lambda^2,r,r')\,s(r',E') \ ,
\end{equation}
where the one-dimensional Green's function $G_{r_h}^{\rm 1D}$ satisfying the boundary conditions of the problem can be obtained by either the method of images:
\begin{equation}
\label{eq:images}
G_{r_h}^{\rm 1D}(\Delta\lambda^2,r,r')=\sum_{k=-\infty}^{\infty}\frac{\Theta(\Delta\lambda^2)}{\sqrt{4\pi\Delta\lambda^2}}\left(\e^{-\frac{(r-r'-2k r_h)^2}{4\Delta\lambda^2}}-\e^{-\frac{(r+r'-2kr_h)^2}{4\Delta\lambda^2}}\right)\ ,
\end{equation}
or by employing Fourier methods:
\begin{eqnarray}
G_{r_h}^{\rm 1D}(\Delta\lambda^2,r,r') &=&\frac{2\,\Theta(\Delta\lambda^2)}{r_h}\sum_{n=1}^{\infty}\sin\left(\frac{n\pi r'}{r_h}\right)\sin\left(\frac{n\pi r}{r_h}\right)\e^{-\frac{n^2\pi^2\Delta\lambda^2}{r_h^2}}\nonumber\\
{} &=&\frac{2\,\Theta(\Delta\eta)}{r_h}\sum_{n=1}^{\infty}\sin\left(\frac{n\pi r'}{r_h}\right)\sin\left(\frac{n\pi r}{r_h}\right)\e^{-n^2\Delta\eta}
\label{eq:fourier}\ .
\end{eqnarray}

In both formulas $\Delta\lambda^2=\lambda^2(E)-\lambda^2(E')$, $\Delta\eta=\eta(E)-\eta(E')$, and $\Theta$ is the Heaviside function.
In a highly non trivial way, it is possible to prove that \eqref{eq:fourier} and \eqref{eq:images} are different representations of the exactly same function.
In appendix \ref{app:analytics} we sketch the derivation of \eqref{eq:fourier} and prove the equivalence between both representations.
The latter formula has not been yet considered in the context of CRE transport in dSphs.
Instead, a solution --henceforth the \old{} formula-- has been systematically employed in the literature. 
This has the same structure of \eqref{eq:masterdiffloss} with the method-of-images representation of the Green's function \eqref{eq:images}, but its $m$-th term differs from ours by a factor $(-1)^k\frac{r}{r-2kr_h}$ (see \eqref{eq:oldmoi}). 

The origin of this formula can be traced back to the appendix A of Ref. \cite{Colafrancesco:2005ji}.
There, the authors attempt to construct the Green's function for the much more general problem of the \emph{3D} heat equation with a the boundary condition at $r=r_h$ by means of the method of images.
Nevertheless, it is possible to verify by inspection, that while the \old{} solution does indeed satisfy the boundary condition, it does not solve the transport equation \eqref{eq:diffloss}.

\subsection{Time-scales hierarchies. Effective descriptions.}
The solution \eqref{eq:masterdiffloss} with either \eqref{eq:images} or \eqref{eq:fourier} is the most general solution to the diffusion-loss problem \eqref{eq:diffloss}.
However, in situations where there is a marked hierarchy between the typical time scales of the problem (diffusion and energy loss), alternative and much simpler, if approximate, answers are available.
Since there are only two relevant time scales in the problem we identify three possible regimes (A, B and C) which are discussed below.

\subsubsection{Regime A. No-diffusion approximation}
In this scenario the energy-loss time scale is much shorter than that of the diffusion. 
Thus $\lambda(E)/r_h\sim\tau_{\rm loss}/\tau_{\rm diff}\ll 1$ and it is appropriate\footnote{Assuming that the DM profile does not have, for instance, any cusps. See, however, the discussion given in section \ref{sec:exact} on this approximation for NFW profiles.} to take the limit $\Delta\lambda^2\to0$ in eq. \eqref{eq:images}. 
The result reads
\begin{equation}
\label{eq:regimeA}
G^A_{r_h}(\Delta\lambda^2,r,r')\to\Theta(\Delta\lambda^2)\left[\delta(r-r')+\delta(r+r')+\delta(r-r'-2r_h)+\ldots\right]\ ,
\end{equation}
where the first term is the only non-vanishing term inside the diffusion zone.
The resulting CRE density is then given by the no-diffusion approximation (henceforth {\bf NDA})
\begin{equation}
n_e(r,E)=\frac{\Theta(r_h-r)}{b(E)}\int_E^\infty\dif E's(r,E')\ ,
\label{eq:rega}
\end{equation}
which, for $r<r_h$, is the solution that is obtained by neglecting the diffusion term in \eqref{eq:diffloss}. 
Note that the \old{} solution also reproduces this limit.

\subsubsection{Regime B. Leading-mode approximation}
In this regime we assume that the time scales of diffusion and energy loss are of the same order so that parametrically $\Delta\lambda^2\sim\lambda^2(E)\sim r_h^2$ and the full solution \eqref{eq:masterdiffloss} must be considered to its fullest.
However, in virtue of the analytical structure of the Fourier-expanded representation of the Green's function \eqref{eq:fourier}, where the term $\e^{-\pi^2\Delta\lambda^2/r_h^2}=\e^{-\Delta\eta}<1$ plays the role of an expansion parameter, we may neglect --to certain degree of accuracy-- all terms with $n>1$ in the series.
Namely,
\begin{equation}
G^B_{r_h}(\Delta\eta,r,r')\to\frac{2\Theta(\Delta\eta)}{r_h}\sin\left(\frac{\pi r'}{r_h}\right)\sin\left(\frac{\pi r}{r_h}\right)\e^{-\Delta\eta}\ .
\end{equation}
Correspondingly, in this leading-mode approximation {\bf (LMA)}, the CRE density reads 
\begin{equation}
n_e(r,E)=\frac2{b(E)r_h}\frac{\sin\left(\pi r/r_h\right)}r\int_E^\infty\dif E'\e^{-[\eta(E)-\eta(E')]}\int_0^{r_h}\dif r'\,r'\sin\left(\frac{\pi r'}{r_h}\right)\,s(r',E') \ .
\label{eq:regb}
\end{equation}

In the following, we will interchangeably use the names \emph{leading-mode}, B${}^{(1)}$ or \emph{RB} approximation when referring to the eq. \eqref{eq:regb}.
Notice that in contrast with the other two regimes, even if the condition $\Delta\lambda^2\sim\lambda^2(E)\sim r_h^2$ would be satisfied (for a given mass and a DM model), the LMA might in some cases be insufficient and higher modes have to be included as well. 
Moreover, since in regimes A and, as we will shortly see, also C, all Fourier modes play a role, the LMA predictions for the emissivities and brightness should {\bf not} be understood as the median of the corresponding regime-A and C predictions, especially if the halos have cusps.
However, when global quantities such as the galaxy-integrated flux densities are considered, it is sensible and quite useful to use the LMA in order to get a quick estimate of the transition between the corresponding regime-A and C computations.
We will numerically assess all these aspects in the sections below.

\subsubsection{Regime C. Rapid-diffusion approximation}
The last possible case is then $\lambda(E)\gg r_h$, i.~e. all the electrons in the galaxy will have enough time to diffuse out all the way to the boundary without losing most of their energy. 
The effective 1D Green's function \eqref{eq:fourier} is given by an asymmetric-triangle function of $r$:
\begin{eqnarray}
\label{eq:regimeC}
G^C_{r_h}(\Delta\lambda^2,r,r')&\to&\delta(\Delta\eta)\sum_{n=1}^\infty\frac2{r_h\,n^2}\sin\left(\frac{n\pi r'}{r_h}\right)\sin\left(\frac{n\pi r}{r_h}\right)\nonumber\\
{}&\to&\delta(\Delta\lambda^2)\left(\frac12(r+r')-\frac12|r-r'|-\frac{rr'}{r_h}\right)\ .
\label{eq:greenrc}
\end{eqnarray}

This expression is obtained by taking the appropriate limit in \eqref{eq:fourier} (see appendix \ref{app:analytics}).
By trivially evaluating the $E'$ integral we obtain the \emph{RC} formula:
 \begin{equation}
n_e(r,E)=\frac1{D(E)}\int\dif r'\frac{r'}{r}\left(\frac12(r+r')-\frac12|r-r'|-\frac{rr'}{r_h}\right)s(r',E)\ ,
\label{eq:regc}
\end{equation}
which is the solution of eq. \eqref{eq:diffloss} in the rapid diffusion approximation {\bf (RDA)}.

\subsection{Summary}
In this section, we reviewed the main assumptions that are made in CRE transport models \eqref{eq:diffloss} which we will scrutinize in the following sections. 
Specifically, we assume that the coefficients in eq. \eqref{eq:diffloss} are given by $D(E)=D_0(E/E_0)^\delta$, where $E_0=1$~GeV, $\delta\in[0.3,0.7]$ and $D_0$ can, at least in principle, be anywhere in the $[10^{22},10^{31}]$~cm${}^2$/s range ($D_0\sim 3\times10^{28}$~cm${}^2$/s is the central value in the MW). 
Similarly, we assume that the all CRE energy losses are due to interactions with the electromagnetic fields: $b(E)\propto u_{\rm EM}E^2$ and that the diffusion-radius parameter $r_h$ is of $\mathcal O(r_\star)$.
The parameter $u_{\rm EM}$ denotes the (constant) ambient electromagnetic energy density in form of radiation and magnetic fields in the dSph. 
The uncertainties of the CRE propagation are thus parametrized by $\tau_0$, $ \delta$, $r_h$ and $B$, where the parameter $\tau_0\equiv r_h^2/D_0$ is more appropriate when scanning the possible transport scenarios.
In particular, this parameter can assume values that are as short as a couple thousands years all the way to $\mathcal O(10^3)$~Gyr.

We assume as well that the particle-injection term can be expressed as $s(r,E)=\mq\,\mr(r)\ms(E)$ and we introduce a couple of benchmark cases for $\mr(r)$ and $\ms(E)$.
By introducing the Syrovatskii variable \eqref{eq:syrovatskii} $\lambda\sim({\rm few~kpc})(E/1{\rm GeV})^{-1/3}$ (for Kolmogorov turbulence and MW-like diffusion constant), the solution \eqref{eq:masterdiffloss} in terms of Green's functions could be found.
There are two representations for these Green's functions: \eqref{eq:images}-\eqref{eq:fourier}.
The first one is constructed by employing the method of images while for the second one, Fourier methods are used.
Finally, we extract from our general solution \eqref{eq:masterdiffloss} the approximated solutions \eqref{eq:rega}-\eqref{eq:regb} and \eqref{eq:regc} for the three regimes that arise once the time-scales hierarchies of the problem are identified. 
In the following we shall exploit the simplifying properties of each one of these solutions.

\section{Universal emissivity profiles and spectra}
\label{sec:universal}
\begin{table}
\centering{
\begin{tabular}{|c|c|c|c|}
\hline
\rule{0pt}{10pt}
\bf Halo model & \bf A \boldmath($\rho(r)$) & \boldmath \bf B${}^{(1)}$  &\bf C\\[2pt]
\hline
\hline
\rule{0pt}{12pt}
PM & $\frac{M_h}{4\pi r^2}\delta(r)$ & $\frac{M_h}{2r_h^2}\frac{\sin(\pi r/r_h)}r$ &  $\frac{\pi M_h(r_h-r)}{4r_h^3r}$ \\[5pt]
\hline
\rule{0pt}{12pt}
Co  & $\rho_h$ & $\frac{2\rho_hr_h}\pi\frac{\sin(\pi r/r_h)}r$ & $\frac{\pi^2\rho_h}6\left(1-\frac{r^2}{r_h^2}\right)$ \\[5pt]
\hline
\rule{0pt}{12pt}
NFW & $\frac{4\rho_sr_s}{r(1+r/r_s)^2}$ & $r_h h_{\rm NFW}^{\rm dec.}\frac{\sin(\pi r/r_h)}r$ & $\frac{2\pi^2\rho_sr_s}{r_h}\left[1-\frac{r}{r_h}-\frac{2r_h}{3r_s}\left(1-\frac{r^2}{r_h^2}\right)+\ldots\right]$ \\[5pt]
\hline
\rule{0pt}{12pt}
mNFW & $\frac{\hat\rho_s\hat r_s^\gamma}{r^\gamma(1+(r/\hat r_s)^\alpha)^{\frac{\beta-\gamma}\alpha}}$ & $r_h h_{\rm mNFW}^{\rm dec.}\frac{\sin(\pi r/r_h)}r$ & solve numerically \\[8pt]
\hline
\rule{0pt}{12pt}
Enst & $\rho_s\e^{-\frac2{\alpha}\left[(r/r_s)^\alpha-1\right]}$ & $r_h h_{\rm Enst}^{\rm dec.}\frac{\sin(\pi r/r_h)}r$ & solve numerically  \\[5pt]
\hline
\rule{0pt}{12pt}
Bkrt & $\frac{\tilde\rho_s}{(1+r/\tilde r_s)(1+r/\tilde r_s)^2}$ & $r_h h_{\rm Bkrt}^{\rm dec.}\frac{\sin(\pi r/r_h)}r$ & $\frac{\pi^2\tilde\rho_s}6\left[1-\frac{r^2}{r_h^2}-\frac{r_h}{2\tilde r_s}\left(1-\frac{r^3}{r_h^3}\right)+\ldots\right]$\\[5pt]
\hline
\end{tabular}}
\caption{\label{tab:halodec} Emissivity halo functions $H_\rho(r)$ for DM decay.}
\end{table} 

By employing the simplified expressions \eqref{eq:rega}, \eqref{eq:regb} and \eqref{eq:regc}, it is straightforward to show that the emissivity (and thus the brightness, flux density, etc.) can be written as the product of a spectral term ($\mx(\nu)$) times a halo term ($H_\mr(r)$):
\begin{equation}
\label{eq:emissmaster}
j_\nu(r) = \frac{\mq}{4\pi}\times H_\mr(r)\times \mx(\nu)\ ,
\end{equation}
where the emissivity halo functions read
\begin{eqnarray}
H_\mr^A(r) &=& \mr(r)\ , \\
\label{eq:rBhalo}
H_\mr^B(r) &=& h_\mr^{(1)}\frac{\sin(\pi r/r_h)}r\ ,\\
H_\mr^C(r) &=& \frac{\pi^2}{r_h^2}\int_0^{r_h}\dif r'\frac{r'}{r}\left(\frac12(r+r')-\frac12|r-r'|-\frac{rr'}{r_h}\right)\mr(r')\ ,
\end{eqnarray}
and the spectral functions are given by
\begin{eqnarray}
\mx_A(\nu) &=& \frac{2\sqrt3e^3B}{m_e}\int\frac{\dif z}z\mf(z)\frac{E(\nu/z)}{b(E(\nu/z))}\int_{E(\nu/z)}^{\infty}\dif E\,S(E)\label{eq:speca}\ , \\
\mx(\nu;\tau_0) &=& \frac{2\sqrt3 e^3 B}{m_e}\int\frac{\dif z}z\mf(z)\frac{E(\nu/z)}{b(E(\nu/z))}\e^{-\eta\left(E(\nu/z)\right)}\int_{E(\nu/z)}^{\infty}\dif E\,S(E)\e^{+\eta(E)}\ ,
\label{eq:specb}\\
\mx_C(\nu) &=&\frac{2\sqrt3e^3B\, r_h^2}{\pi^2m_e}\int\frac{\dif z}z\mf(z)\frac{E(\nu/z)S(E(\nu/z))}{D(E(\nu/z))}\ ,
\label{eq:specc}
\end{eqnarray}
where we have implicitly assumed that electrons and positrons contribute equally. 
Expressions \eqref{eq:speca} and \eqref{eq:specc} can also be obtained by taking the no-diffusion and the rapid-diffusion limits on \eqref{eq:specb}, respectively.
More specifically, the spectral function \eqref{eq:specb} tends to \eqref{eq:speca} in the $\lambda^2(E)\ll r_h^2$ ($\tau_0\to\infty$) limit (independently of the functional form of $\ms(E)$).
Less trivially, the spectrum \eqref{eq:specb} is also such that $\mx\to\mx_C$ when $\lambda^2(E)\gg r_h^2$ ($\tau_0\to0$). 
See appendix \ref{app:analytics}.
These arguments show that eq. \eqref{eq:specb} comprises all possible cases and we may omit putting a label (regime-B) on it.
Instead, we explicitly add the parameter $\tau_0$ into its argument in order to emphasize this universality.

\begin{table}
\centering{
\begin{tabular}{|c|c|c|c|}
\hline
\rule{0pt}{10pt}
\bf Halo model & \bf A &\boldmath \bf B${}^{(1)}$  &\bf C\\[2pt]
\hline
\hline
\rule{0pt}{15pt}
PM & $\frac{K_{\rm pm}}{4\pi r^2}\delta(r)$ & $\frac{K_{\rm pm}}{2r_h^2}\frac{\sin(\pi r/r_h)}r$ &  $\frac{\pi K_{\rm pm}(r_h-r)}{4r_h^3r}$ \\[5pt]
\hline
\rule{0pt}{15pt}
Co  & $\rho_h^2$ & $\frac{2r_h\rho_h^2}{\pi}\frac{\sin(\pi r/r_h)}r$ & $\frac{\pi^2\rho_h^2}6\left(1-\frac{r^2}{r_h^2}\right)$ \\[5pt]
\hline
\rule{0pt}{15pt}
NFW & $\frac{16\rho_s^2r_s^2}{r^2(1+r/r_s)^4}$ & $h_{\rm NFW}^{\rm ann.}\frac{\sin(\pi r/r_h)}r$ & $\frac{16\pi^2\rho_s^2r_s^2}{r_h^2}\left[\ln\frac{r_h}r-2\frac{r_h}{r_s}\left(1-\frac{r}{r_h}\right)+\ldots\right]$ \\[5pt]
\hline
\rule{0pt}{15pt}
mNFW & $\frac{\hat\rho_s^2\hat r_s^{2\gamma}}{r^{2\gamma}(1+(r/\hat r_s)^\alpha)^{\frac{2(\beta-\gamma)}\alpha}}$ & $h_{\rm mNFW}^{\rm ann.}\frac{\sin(\pi r/r_h)}r$ & solve numerically \\[8pt]
\hline
\rule{0pt}{15pt}
Enst & $\rho_s^2\e^{-\frac4{\alpha}\left[(r/r_s)^\alpha-1\right]}$ & $h_{\rm Enst}^{\rm ann.}\frac{\sin(\pi r/r_h)}r$ & solve numerically  \\[5pt]
\hline
\rule{0pt}{15pt}
Bkrt & $\frac{\tilde\rho_s^2}{(1+r/\tilde r_s)^2(1+r/\tilde r_s)^4}$ & $r_h h_{\rm Bkrt}^{\rm ann.}\frac{\sin(\pi r/r_h)}r$ & $\frac{\pi^2\tilde\rho_s^2}6\left[1-\frac{r^2}{r_h^2}-\frac{r_h}{\tilde r_s}\left(1-\frac{r^3}{r_h^3}\right)+\ldots\right]$\\[5pt]
\hline
\end{tabular}}
\caption{\label{tab:haloann} Emissivity halo functions $H_{\rho^2}(r)$ for DM annihilation.}
\end{table} 

The factorization property of \eqref{eq:emissmaster} is reminiscent of the prompt gamma-ray flux prediction from DM annihilation/decay.
In fact, as we shall shortly observe, the flux-density halo functions in our regime A and the corresponding halo functions for gamma rays  (better known as $J$ and $D$ factors) are exactly the same. 

In tables \ref{tab:halodec} and \ref{tab:haloann} we include analytical formulas for the emissivity halo functions for various DM profiles\footnote{The corresponding brightness (flux-density) halo functions (factors) are given in tables \ref{tab:bhalodec} and \ref{tab:bhaloann} (\ref{tab:fdhalo}).}. 
In the leading-mode approximation, we obtain a very interesting result: all halo functions have identical shapes (independent of $\mr(r)$).
All the $\mr$ dependence is instead captured in the halo \emph{factors} $h_\mr^{(1)}$ which are tabulated in Table \ref{tab:hfactor}. 
The meaning of the upper index ``(1)'' on the halo factors $h_\mr^{(1)}$ in \eqref{eq:rBhalo} will become clear in the next section. 
Note that, in obtaining the entries of tables \ref{tab:halodec}-\ref{tab:haloann}, we assume that the scale radii $r_s$ are always much larger than $r_h$ and for simplicity, we Taylor-expand our formulas on the $r_h/r_s$ ratio.
As it is conventional, the scale radius $r_s$ is defined by the condition $\left.\frac{r}{\rho}\frac{\dif\rho}{\dif r}\right|_{r=r_s}=-2$ and $\rho_s$ is such that $\rho(r_s)=\rho_s$\footnote{For the modified-NFW and Burkert profiles we introduced the parameter $\hat r_s$ and $\tilde r_s$ respectively. These are related with the corresponding scale radii as follows $\hat r_s =\left(\frac{\beta-2}{2-\gamma}\right)^{1/\alpha}r_s$ and $\tilde r_s=\frac{r_s}{(1-\sqrt{78/81})^{1/3}+(1+\sqrt{78/81})^{1/3}}\approx0.65728\,r_s$.}.

The spectral functions, on the other hand, need to be computed numerically even in the monochromatic electron injection case: $\ms(E)=\delta(E-E')$. 
The only exception is in the regime C, where $\mx_C^{\,e^+e^-}=\frac{4\sqrt3e^3B\,\tau_0}{\pi^2m_e(E'/E_0)^\delta}\times \mf\left(\frac{2\pi m_e^3\nu}{3eBE'^2}\right)$ has a manifestly closed analytical form in terms of the Macdonald functions shown in \eqref{eq:fav}.
Note also that out of the four transport-model parameters ($\tau_0$, $r_h$, $\delta$ and $B$), the halo functions can only depend on $r_h$ while on the other hand, the spectral functions are $r_h$-independent.
Moreover, in the regime-A situation the result is --as it should be-- independent of $\tau_0$ and $\delta$.  

\subsection{Exact treatment}
Using the Fourier representation \eqref{eq:fourier} of the Green's function we can easily obtain the exact expression for the emissivity.
This is given by an infinite sum of terms with halo and spectrum functions multiplied together:
\begin{equation}
\label{eq:emissexact}
j_\nu(r) = \frac{\mq}{4\pi}\sum_{n=1}^\infty h_\mr^{(n)}\frac{\sin\frac{ n\pi r}{r_h}}r\mx\left(\nu,\frac{\tau_0}{n^2}\right)\ ,
\end{equation}
where the $n$-th order halo factor $h^{(n)}_\mr$ is defined as 
\begin{equation}
\label{eq:nthhalo}
h_\mr^{(n)}=\frac{2}{r_h}\int_0^{r_h}\dif r \, r\,\sin\frac{n\pi r}{r_h}\mr(r)\ .
\end{equation}

In table \ref{tab:hfactor}, formulas for the factors are shown for a number of DM halo models. 
Using the properties of the spectral function \eqref{eq:specb}, we can immediately relate these factors with the regime A-C halo functions. 
Concretely, for very large $\tau_0$ we have seen that $\mx\left(\nu,\frac{\tau_0}{n^2}\right)\to\mx_A(\nu)$ (independent of $n$). 
Analogously, the spectral function has an inverse-$n^2$ dependence ($\mx\left(\nu,\frac{\tau_0}{n^2}\right)\to\mx_C(\nu,\tau_0)/n^2$) for small enough $\tau_0$. 
Thus, 
\begin{equation}
\label{eq:hfourier}
H_\mr^A(r)=\sum_{n=1}^\infty h_\mr^{(n)}\frac{\sin\frac{ n\pi r}{r_h}}r\quad ,\quad H_\mr^C(r)=\sum_{n=1}^\infty \frac{h_\mr^{(n)}}{n^2}\frac{\sin\frac{ n\pi r}{r_h}}r\ .
\end{equation}
We verified that all entries in our tables satisfy these conditions.
Expressions \eqref{eq:hfourier} are also useful for understanding the transition between the different regimes.
We will explore this with more detail in the sections below.

\begin{table}
\centering{
\begin{tabular}{|c|c|c|}
\hline
\rule{0pt}{10pt}
{} & 
\bf Annihilation  &\bf Decay \\[2pt]
\hline
\hline
\rule{0pt}{15pt}
PM & 
$\frac{nK_{\rm pm}}{2r_h^2}$ &  $\frac{nM_h}{2r_h^2}$ \\[5pt]
\hline
\rule{0pt}{15pt}
Co  & 
$\frac{2\,r_h}{n\pi}(-1)^{n-1}\rho_h^2$ & $\frac{2\,r_h}{n\pi}(-1)^{n-1}\rho_h$ \\[5pt]
\hline
\rule{0pt}{15pt}
NFW & 
\begin{tabular}{c}
$32\left({\rm si}(n\pi)-\frac8{n\pi}\frac{r_h}{r_s}+\ldots\right)\frac{r_s^2\rho_s^2}{r_h}$, odd $n$\\
$32\left({\rm si}(n\pi)+\mathcal O\left(\frac{r_h^2}{r_s^2}\right)\right)\frac{r_s^2\rho_s^2}{r_h}$, even $n$
\end{tabular}
& \begin{tabular}{c}
$\frac{16}{n\pi}\left(1-\frac{r_h}{r_s}+\ldots\right)r_s\rho_s$, odd $n$\\
$\frac{16}{n\pi}\left(0+\frac{r_h}{r_s}+\ldots\right)r_s\rho_s$, even $n$\\ 
\end{tabular}
\\[5pt]
\hline
\rule{0pt}{15pt}
Bkrt & 
\begin{tabular}{c}
$\frac{2\,r_h}{n\pi}\left(1-2\left(1-\frac{4}{n^2\pi^2}\right)\frac{r_h}{\tilde r_s}\right.$\\[5pt]
$+\ldots\Big{)}\tilde\rho_s^2$, odd $n$ \\[5pt]
$\frac{2\,r_h}{n\pi}\left(-1+2\frac{r_h}{\tilde r_s}+\ldots\right)\tilde\rho_s^2$, even $n$
\end{tabular}
& 
\begin{tabular}{c}
$\frac{2\,r_h}{n\pi}\left(1-\left(1-\frac4{n^2\pi^2}\right)\frac{r_h}{\tilde r_s}\right.$\\[5pt]
$+\ldots\Big{)}\tilde\rho_s$, odd $n$\\[5pt]
$\frac{2\,r_h}{n\pi}\left(-1+\frac{r_h}{\tilde r_s}+\ldots\right)\tilde\rho_s$, even $n$
\end{tabular}
\\[5pt]
\hline
\end{tabular}
}
\caption{\label{tab:hfactor} $n$-th order halo factors for several benchmark mass models (expanded in the $r_h/r_s\ll 1$ ratio). si($x$) is the sine integral function.}
\end{table}

\section{Numerical strategies and results}
\label{sec:numerics}
\subsection{Halo functions}
\begin{figure}[t]
\includegraphics[width=\linewidth]{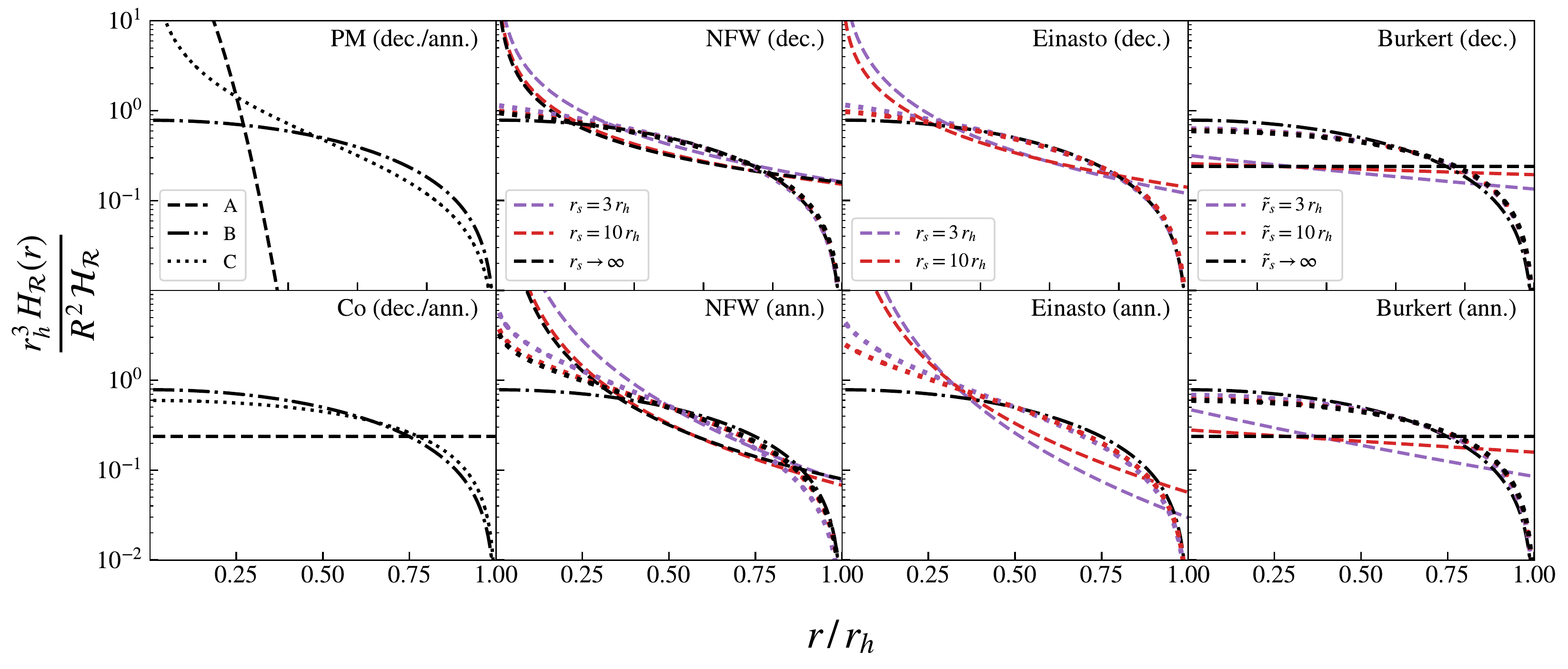}
\caption{\label{fig:halos} Normalized emissivity halo functions for several DM-mass profiles in the RA (solid), RB (dotted) and RC (dashed) regimes. For NFW, Einasto and Burkert profiles the variations of the scale-radius parameter are also shown: $r_s=3\times r_h$ (purple), $r_s=10\times r_h$ (red) and $r_s\to\infty$ (black). In the first panel (PM model), we use the Gaussian approximation for the 3D Dirac delta: $1/(4\pi r^2)\times\delta(r)\to1/(2\pi\sigma_0^2)^{3/2}\times\e^{\,-r^2/(2\sigma_0^2)}$. The flux-density halo factors ($\mh_\mr$) are defined and computed in the appendix.}
\end{figure}

Fig. \ref{fig:halos} shows the radial dependence of the regime-A, -B and -C emissivity halo functions for our DM-profile parametrizations ``PM'' and ``Co'' for both DM annihilation and decay as these are, up to normalization, identical. 
We also include there the halo functions that result from considering the NFW, Einasto and Burkert parametrizations for DM annihilation or decay.
In appendix \ref{app:brighthalos}, analogue results for the brightness (in terms of angular variables) are also plotted.

These results are obtained by evaluating the formulas shown in tables \ref{tab:halodec} and \ref{tab:haloann} and are then divided by their volume integral over the whole galaxy.
In other words, we divide the expressions by the flux-density halo factors \eqref{eq:fdhalo} displayed in table \ref{tab:fdhalo} and the ratio $R^2/r_h^2$. 
In the case of the Einasto profile, a couple of (simple) numerical integrations are necessary.

A couple interesting things happen here. 
First, bear in mind that the volume integrals of all curves in fig. \ref{fig:halos} are, by construction, normalized to one.
In particular, small-$r$ behaviour of the halo functions in the different regimes does not need to follow the (intuitively expected) ``A $>$ B $>$ C'' hierarchy.  
Second, we observe that in going from the NDA (RA) to the LMA (RB) approximations, large qualitative differences appear. 
However, the regime-C emissivity halo functions do not depart significantly from those in the regime B, especially in the case of decaying DM with an NFW profile.
This is a direct consequence of the rapid Fourier convergence of the effective Green's function \eqref{eq:greenrc} in the RDA.
Specifically, the $n$-th term of the series is suppressed by a factor $1/n^2$ (see first line of \eqref{eq:greenrc}).
In the case of decaying DM in a NFW-like halo the similarity between the two regimes is particularly acute.
This is because the $n=2$ term (in the $r_s\ll r_h$ limit) vanishes accidentally (see corresponding entry in table \ref{tab:hfactor}).

In order to make the shape comparisons quantitatively, we computed half-flux radii (HFR) for all cases (decaying and annihilating DM in several halo-shape parametrizations) and regimes. 
These are tabulated in table \ref{tab:fdhalo}.
We observe that, for instance, the HFRs in the regime C are numerically close to their corresponding regime-B HFRs ($\simeq 0.61\,r_h$).
More specifically, in the most extreme ``point-mass profile'' case --which is such that the disagreement between the B and C halo functions is maximized-- a modest $\sim$ 20\% difference is found.
Instead, the difference between the regime-A and the regime-B (or C) HFRs for this case amounts to 100\%.
Conversely, the regime-C HFR is only $\sim1$\% larger than its corresponding regime-B HFR in a NFW halo when DM decay is the process of interest.

\subsection{Spectra}
\begin{figure}[t]
\includegraphics[width=\linewidth]{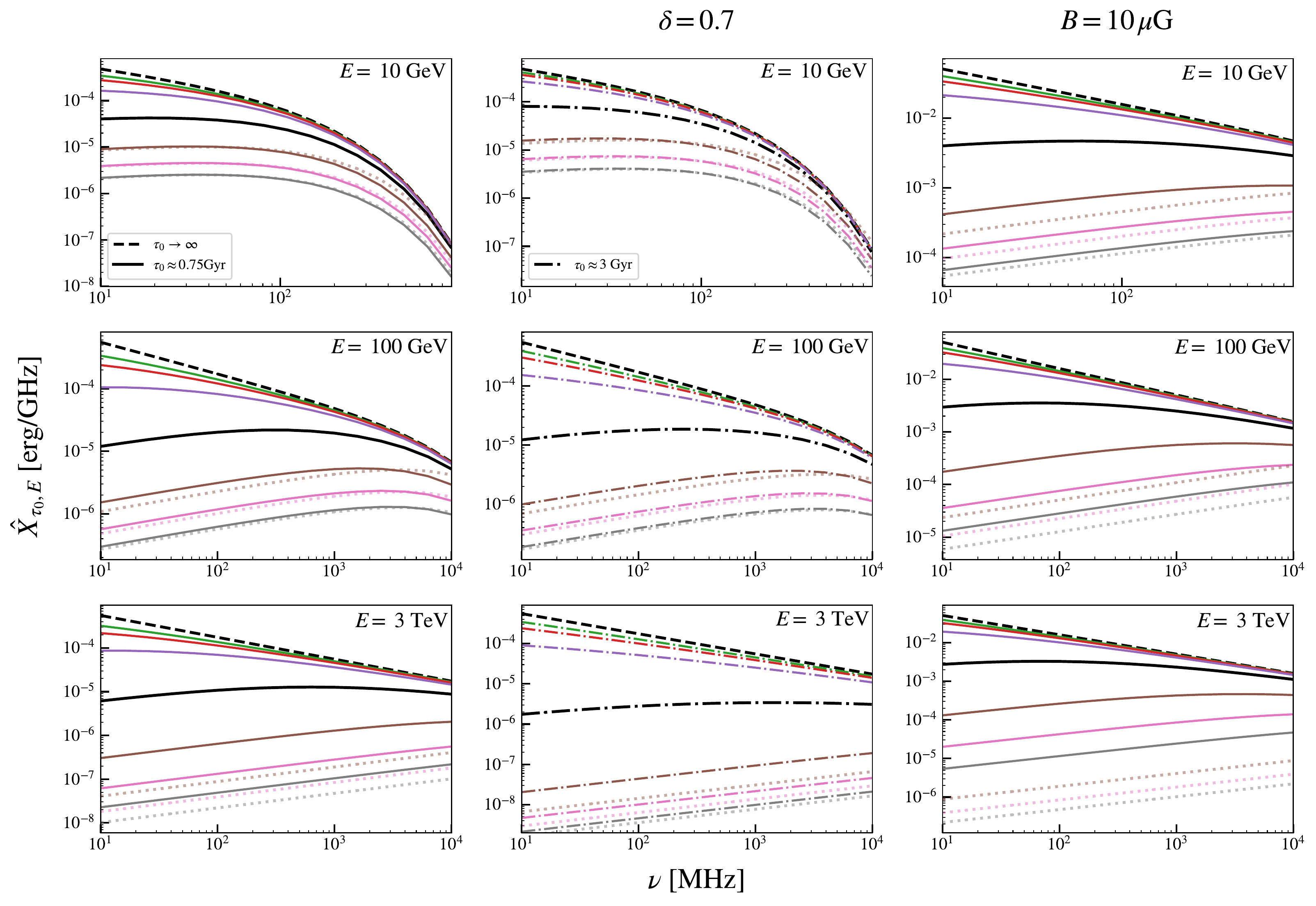}
\caption{\label{fig:eetree} Spectral functions $\hat\mx_{\tau_0,E}(\nu)$ (tree-level $e^+e^-$ process) for several transport-model parameters ($\delta$ and $B$) and values for $E$. Solid lines obtained using \eqref{eq:specb} with the central values $\tau_0\simeq0.75$~Gyr (solid black) and 3~Gyr (dot-dashed black). Color-coded curves correspond to 16$\times$ (green), 9$\times$ (red), 4$\times$ (purple), (1/4)$\times$ (brown), (1/9)$\times$ (pink) and (1/16)$\times$~0.75~Gyr (3~Gyr) (in gray).
Dashed black lines show the RA \eqref{eq:speca} spectra (NDA) while RC \eqref{eq:specc} (RDA) results are plotted using dotted lines respecting the color coding just described.}
\end{figure}

The resulting spectra for the ``vanilla'' $e^+e^-$ models are shown in fig. \ref{fig:eetree}.
Before discussing the figure, a few words on our numerical methods for the computation of \eqref{eq:specb} are in order.
First of all, as already noted, the propagation parameter $\tau_0$ plays a leading role in the determination of the spectra \eqref{eq:specb}.
This parameter is very uncertain and the evaluation of the function $\mx(\nu;\tau_0)$ for a multi-scale range of $\tau_0$ values is therefore an unavoidable task.
Fortunately, these scans can be reused in e.~g. improving the accuracy of our predictions.  
More specifically, a scan with a sufficiently dense $\tau_0$-grid can be used in order to determine the leading and sub-leading spectral functions $\mx(\nu;\tau_0)$, $\mx(\nu;\tau_0/4)$, $\mx(\nu;\tau_0/9)$, etc. 
By plugging these with the relevant halo factors from table \ref{tab:hfactor} into \eqref{eq:emissexact}, we can obtain more and more accurate predictions.
Furthermore, in order to make our computations for generic DM models even more efficient, we also exploit another analytical property of eq. \eqref{eq:specb}.
Namely, the spectral function $\mx(\nu;\tau_0)$ in the most generic case can be written as
\begin{equation}
\label{eq:genericspec}
\mx(\nu;\tau_0)=\int\dif E\,\ms(E)\hat\mx_{\tau_0,E}(\nu)\ ,
\end{equation}
where we define here $\hat\mx_{\tau_0,E}(\nu)$: the spectral function that results from monochromatic $e^+e^-$ injections into the halo ($S_E(E')=\delta(E-E')$).
In particular, $E=\mchi$ ($E=\mchi/2$) if the relevant process is DM annihilation (decay).
Concretely,
\begin{equation}
\hat\mx_{\tau_0,E}(\nu)= \frac{2\sqrt3 e^3 B}{m_e}\int_{\frac{2\pi m_e^3\nu}{3eBE^2}}^\infty\frac{\dif z}z\mf(z)\frac{E(\nu/z)}{b(E(\nu/z))}\e^{-[\eta\left(E(\nu/z)\right)-\eta(E)]}\ .
\end{equation}

As with the $\tau_0$ parameter, a densely spaced $E$ scan of $\hat\mx_{\tau_0,E}$ will be necessary not only in order to scan over the DM mass in the vanilla $e^+e^-$ models, but in virtue of \eqref{eq:genericspec}, these evaluations will enable us to determine the spectra in the most generic case.
Fortunately, this is a not very expensive computational task.
In particular, by employing the right numerical strategies, all our spectra (for generic models) can be reproduced in a matter of seconds.

Fig. \ref{fig:eetree} shows the $\hat\mx_{\tau_0,E}(\nu)$ spectra for several values of the transport-model parameters. 
In particular, for the spectra shown on the left column we assumed that $B=0.1~\mu$G, $\delta=0.3$.
Unless stated differently, this will be henceforth our standard choice, as well as the benchmark frequency of 150~MHz.
We also included their corresponding RA and RC spectra using \eqref{eq:speca} and \eqref{eq:specc} respectively.
As expected, for a given magnetic field and $E$, the RA formula gives the largest possible spectrum.
This curve exhibits a power-law like ($\propto 1/\sqrt\nu$) behaviour for $\nu\lesssim \mathcal O(eBE^2/m_e^3)$.
For larger frequencies the spectrum is of course exponentially suppressed.

In the opposite limit, on the other hand, we see that the RC formula reproduces the exact result very well 
for small enough $\tau_0$.
In this limit all spectra feature the $X\propto\nu^{1/3}$ (for $\nu\lesssim \mathcal O(eBE^2/m_e^3)$) as in the familiar single-emitter synchrotron power formula.
We remind the reader that for regime C, CREs do not have time to lose their energies and therefore, the spectrum is well described by the single-emitter formula. 
We also observe that, as expected, the heavier the DM particle and the stronger the magnetic field, the smaller the values of $\tau_0$ around which the RC formula becomes appropriate. 

\begin{figure}[t!]
\includegraphics[width=\linewidth]{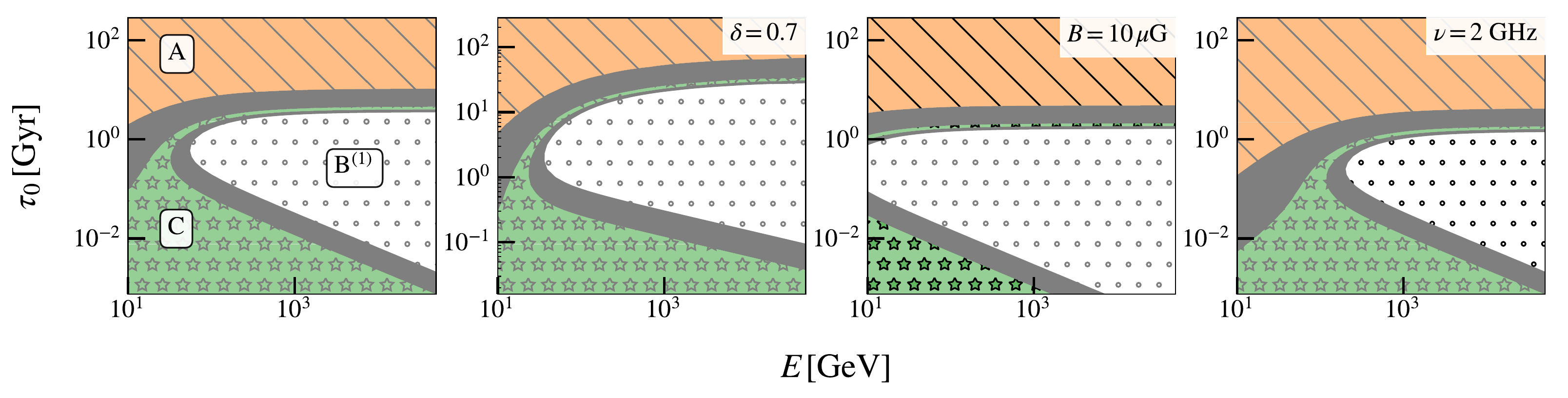}
\includegraphics[width=\linewidth]{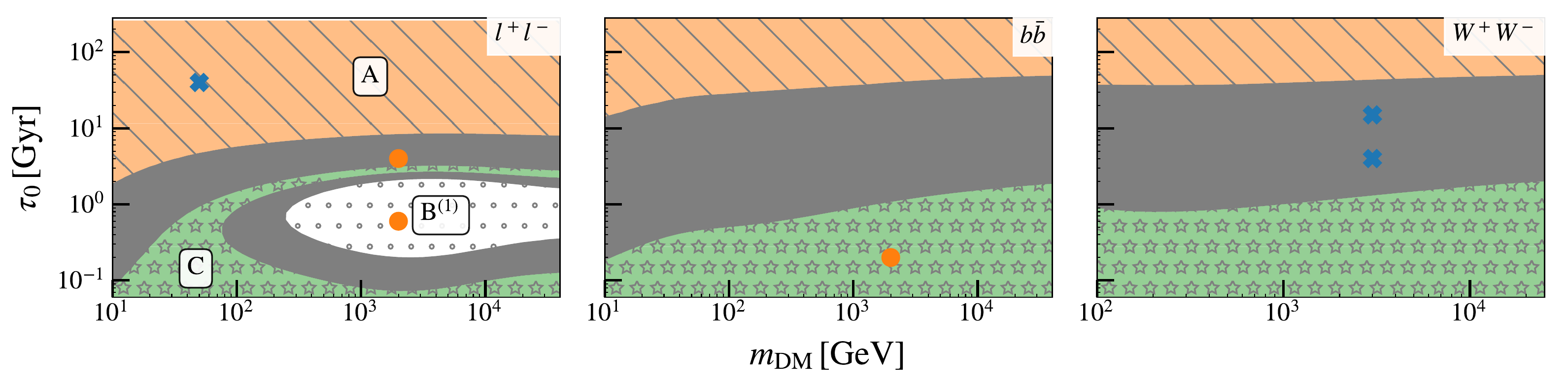}
\caption{\label{fig:contours} Upper panels: regions in the $(\tau_0,E)$-plane (with $B$ and $\delta$ fixed) where the RA (orange, diagonal hatch pattern), RB (white, dotted pattern) and RC (green, starred) formulas for $\hat\mx_{\tau_0,E}$ are valid. 
Mixed regimes are indicated in gray color and no hatches.
The default values for the other relevant parameters are $\nu=150$~MHz, $B=0.1\,\mu$G and $\delta=0.3$. 
Changes on any of these are indicated on the upper-right corner of the plots.
Lower panels: Same as upper ones but for the complete $\mx(\nu,\tau_0)$ spectra and for several choices of $\ms(E)$ ($l^+l^-$, $W^+W^-$ and $b\bar b$) for DM annihilation (in case of DM decay, the abscissa has to be rescaled by a factor of two: $\mchi\to\mchi/2$ ).
The blue crosses and red blobs indicate the benchmark points in the parameter space that are considered in section \ref{sec:exact} and appendix \ref{app:bench} respectively.}
\end{figure}

In order to gain visual understanding of these points, we show in fig. \ref{fig:contours} the ranges of validity in the $(\tau_0,E)$ plane of the A, B and C formulas.
Specifically, we compute at every point the quantity $f_2=\hat\mx_{\frac{\tau_0}4,E}/\hat\mx_{\tau_0,E}$, i.~e. the ratio of the 2th- to the 1st-order spectral function (see \eqref{eq:emissexact}) and we then set threshold values for every approximation: $f_2>1/2$ for RA,  $\frac14\times0.9<f_2<\frac14\times1.1$ for RC and  $f_2<20\%$ in the RB case. 
These conditions are constructed by noting that in the pure RA situation $f_2\to1$, while $f_2\to1/2^2=1/4$ if the RC formula were exact. 
The latter scaling can be verified by a simple inspection of the Fourier-mode decomposition of the RC Green's function \eqref{eq:greenrc}.

We caution, however, that alternative and probably more robust definitions for the regions in the figure are possible.
A drawback of the simplicity of our prescription is that, for instance, the thin green-starred funnel in the mixed A-B regime has been misidentified as C. 
A more robust definition of the regions could, e.~g. consider power corrections of the RA and RC effective Green's functions\footnote{For instance, by including the correction $\delta G_{r_h}^C=-\delta'(\Delta\eta)\sum\frac{2}{r_h n^4}\sin(n\pi r'/r_h)\sin(n\pi r/r_h)$ of the regime-C Green's function \eqref{eq:greenrc} on the computation and comparing the result with the uncorrected one.} or incorporate the $n$-dependence of the halo factors of table \ref{tab:hfactor}.
We content ourselves with this simplified approach as it proves quite effective.

This said, we move on to discuss the main features of fig \ref{fig:contours}.
In particular, we observe rather well-defined power-law like boundaries between the regimes A and B, but also between B and C.
In order to get a feeling about these, note that the computation is characterized by two energy (and thus, Syrovatskii-length) scales: $E$ and $E(\nu)$. 
Assuming that $E\gg E(\nu)$, the computation of the mixed A-B regime spectrum will be dominated by terms satisfying $\lambda(E(\nu))\sim r_h$ or, by using \eqref{eq:syrint}, $\tau_0\propto(1-\delta)^{-1}(B/\nu)^{1-\delta}/[1+(B/3.13\mu{\rm G})^2]$, which independent of $E$.
The formula explains why the A-B boundary moves up $\delta$ and $\nu$ are increased. 
The $B$-field dependence of the boundary is less trivial, though. 
Above the peculiar $\sim3~\mu$G $B$-field magnitude\footnote{The magnetic and CMB energy densities are equal for $B\approx3\mu$~G.} the boundary moves down as observed in the plot.

The lower transition between B and C is instead characterized by the highest energies of the problem.
This is because in the RDA (small $\tau_0$) electrons do not have time to loss their initial energies.
Thus, $\lambda(E)\sim r_h$ and $\tau\propto(1-\delta)^{-1}E^{-(1-\delta)}/[1+(B/3.13\mu{\rm G})^2]$.
In particular, the boundary is $\nu$-independent as can be verified by comparing the first and fourth plots in the upper panel of the figure.
In the $E(\nu)\gg E$ limit the leading-mode approximation is not appropriate for any value of $\tau_0$.
Therefore, in the crossover region between RA and RC, the full solution \eqref{eq:emissexact} might be necessary.
We also observe, as intuitively expected, that when the magnetic field is increased the RB boundary moves to the left (lower energies), while the opposite occurs if we increase the frequency.

In the lower panel of the figure we also include the boundaries between the regimes for specific models for $\ms(E)$. 
In particular, for the $W^+W^-$ and $b\bar b$ models we do not expect that the B${}^{(1)}$ approximation is accurate for any combination of parameters. 
This is because most of the electrons that are created by the annihilation/decay of DM into these channels have much lower energies as $E\sim\mathcal O(\mchi)$. 
The opposite occurs --as expected-- when the primary process involves leptonic states, such as the $e^+e^-$ final state or the charged-lepton model $l^+l^-=(e^+e^-+\mu^+\mu^-+\tau^+\tau^-)/3$ that we introduced in the figure.

\subsection{Numerical tests} 
\label{sec:exact}
Armed with the numerical strategies discussed in previous sections, we can test the performance of our approximated formula \eqref{eq:emissmaster} across the different characteristic regimes of the problem.
Specifically, as mentioned above, we have devised a series of methods that are very efficient in evaluating the (exact) formula \eqref{eq:emissexact} which will allow us to test the accuracy of the RA--RC formulas.

\begin{figure}[h!]
\centering{
\includegraphics[width=.5\linewidth]{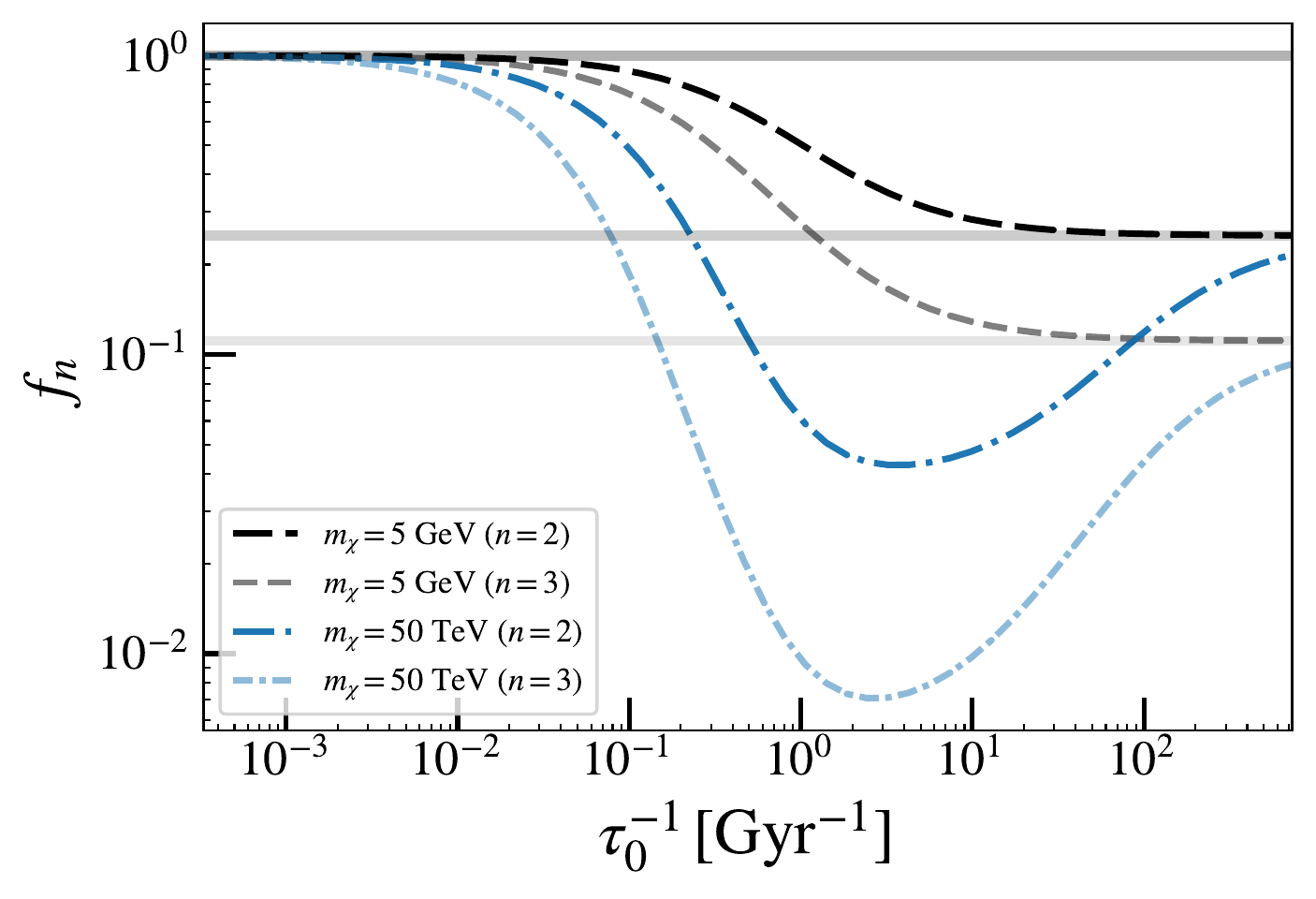}}
\caption{\label{fig:subleadratios} $f_n(\tau_0,E)$ ratios for different values of $E$ as a function of $\tau_0$. 
The RA ($f_n=1$ thick gray line) and RC ($f_n=1/n^2$ thinner gray lines) limits are asymptotically reached in all cases.
The RB limit $f_{n\neq1}\to0$ is instead reached for larger values of $E$ (manifested as valley-shaped features in the blue dot-dashed curves).
We assume that $B=0.1\,\mu$G, $\delta=0.3$ and $\nu=150$~MHz.}
\end{figure}

We begin by discussing the effect of contributions of order higher than $n=1$ and define the spectrum ratios $f_n(\tau_0,E)=\hat\mx_{\tau_0/n^2,E}/\hat\mx_{\tau_0,E}$.
By definition, these ratios are such that $f_n\to1$,  $f_n\to0$ and $f_n\to1/n^2$ for the RA, RB and RC formulas to be valid respectively.
We thus expect, that for fixed $E$ and large enough $\tau_0$, the RA condition is met while in the opposite limit $f_n\to1/n^2$ asymptotically. 
For intermediate values of $\tau_0$, the ratio $f_n$ may become smaller than $1/n^2$ and if that is the case, then the RB formula is appropriate there.   
This is precisely what we find in fig. \ref{fig:subleadratios} where, using our benchmark parameter set and frequency, we consider these ratios for the first two harmonics $n=2$ and $n=3$.
As already observed in fig. \ref{fig:contours}, the validity of the LMA is only achieved when $E$ is large enough.
The figure provides us with a strong confidence on the numerical implementation of our results and we may thus center our attention on the discussion of the exact results in mixed regimes.

For definiteness, we will consider three benchmark points (blue crosses) shown on the lower-right and left panels of fig. \ref{fig:contours}.
Namely, (1) the $l^+l^-$ model for annihilating (decaying) DM with mass of 50~GeV (100~GeV) in a NFW halo and an almost purely RA configuration ($\tau_0=40$~Gyrs with $B=0.1\,\mu$G, $\delta=0.3$ and $\nu=150$~MHz).
A second (2) and third (3) benchmark points are also considered: $W^+W^-$ model with the same parameters but with the (much larger) mass of 3~TeV (in the DM annihilation hypothesis) and $\tau_0=4$~Gyr and 15~Gyr.

The first benchmark point has been chosen in such a way that the quality of the RA formula can be assessed, while the second and third ones will enable us to understand the transition between the RA and RC formulas.
Note that, in contrast with the transition between formula A ($f_n\simeq1$) and either B or C, the B-to-C transition is expected to be very smooth since for both cases, the ratio $f_n$ decays as a function of $n$ rather quickly.
In appendix \ref{app:bench} we also include further benchmark points assessing the RA-RB transition and the accuracy of the RC formula.
These are indicated as red circles in the lower-right and center panels of the figure.

\begin{figure}[t!]
\includegraphics[width=\linewidth]{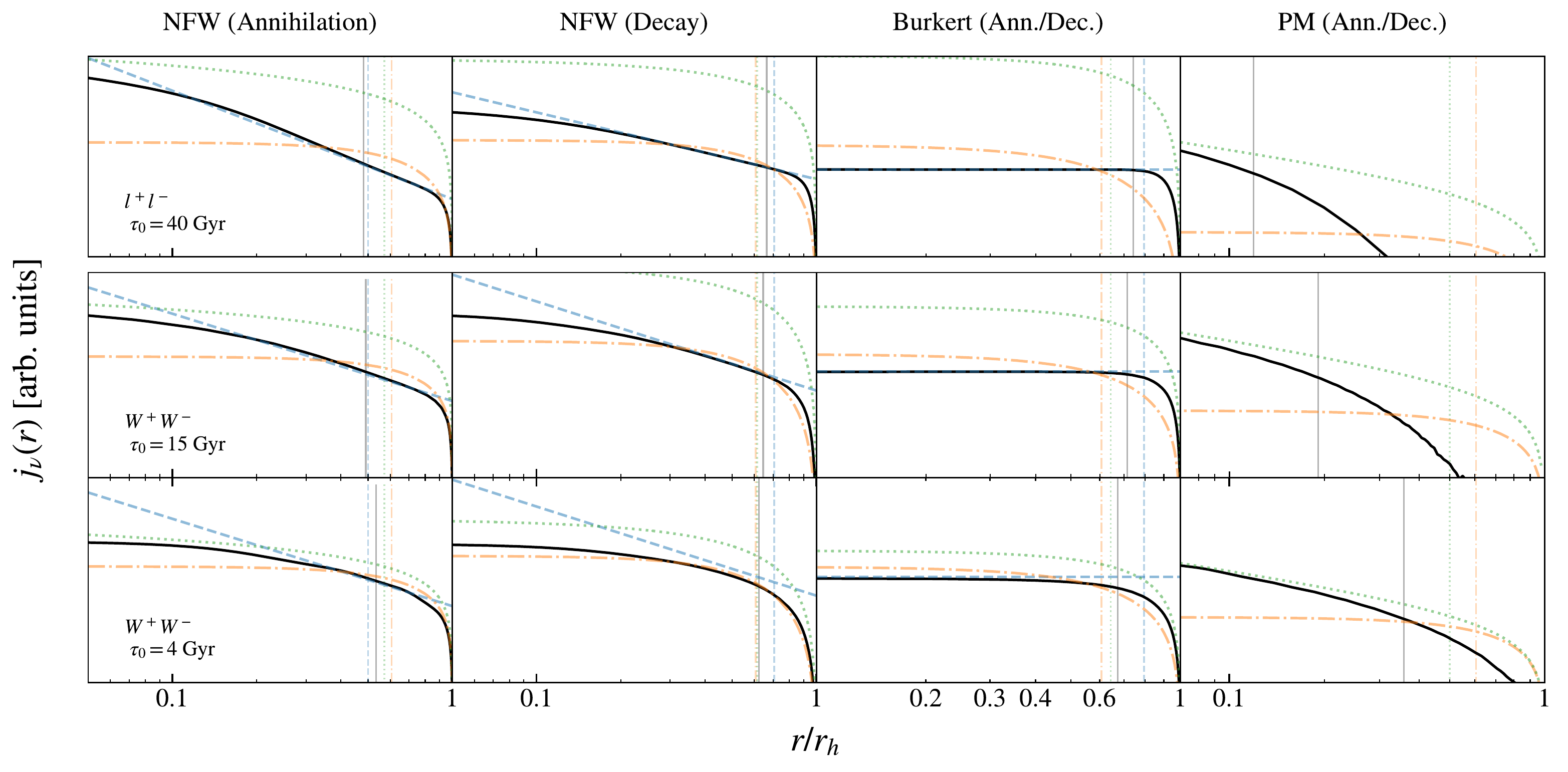}
\caption{\label{fig:exact} Emissivity profiles for the benchmark scenarios indicated with blue crosses in fig \ref{fig:contours}. 
Concretely, a leptophilic model with $\mchi=50$~GeV (ann.) in the first row and a wino-like model with $\mchi=3$~TeV for the second and third rows.
Exact computations in arbitrary units with logarithmic abscissas and ordinates are shown in black. 
The corresponding RA (light dashed blue), RB${}^{(1)}$ (light dot-dashed orange) and RC (dotted green) computations are also shown together with their associated half-flux radii (HFRs) represented as vertical lines.
In the case of the point-like mass model the RA prediction is proportional to a Dirac delta which we omit plotting.}
\end{figure}

Fig. \ref{fig:exact} shows the results of our comparisons.
In all cases, we include the exact \eqref{eq:emissexact} and the approximated \eqref{eq:emissmaster} (with \eqref{eq:speca}-\eqref{eq:specc}) computations of the emissivity for several halo models and within the benchmark parameter sets introduced above.
In order to quantify the morphological resemblance between the different predictions, we also show their corresponding HFRs (vertical lines). 
As evident from the figure, this quantity does a remarkably good job in pinning down the similarities/differences between the various emissivity-function shapes that we considered.

Our results prove --at the numerical level-- several claims made before. 
Indeed, the upper panels show that for the leptophilic model with a 50~GeV DM-mass, our effective RA description agrees very well with the exact results regardless of the inner structure (cusped or cored) of the DM density.
The small-$r$ behaviour of the solutions merits some discussion, though.
First, we see that for the cored profile (Burkert) the emissivity mirrors the functional behaviour of the ``parent'' DM profile while the B and C computations overestimate it when $r$ goes to zero.
Conversely, in the case that the DM density has a cusp, we verified that independent of $\tau_0$, the small-$r$ behaviour the emissivity tends to our RC result. 
In particular, in the NFW (annihilation) and point-mass halo models we verified that, for sufficiently small $r$, the emissivities follow the $j_\nu\propto\ln(1/r)$ and $j_\nu\propto r^{-1}$ radial dependences respectively and futhermore, the correct normalization is also reproduced by our RC formula.
In the case of decaying DM (NFW), instead, the emissivity as a function of $r$ develops a core that is not accounted for in the RA formula, which has the familiar $\propto 1/r$ cusp (see table \ref{tab:halodec}).

Physically, this curious small-$r$ behaviour is just a manifestation of the fact that the typical diffusion time scale at the vicinity of $r=0$ ($\sim r^2/D(E)$) is much smaller than the typical energy-loss time scale. 
Thus, for sufficiently small $r$, the otherwise dominant energy-loss term can be neglected and the RC description is (locally) appropriate.
For cored profiles, the smallest radial scale is $r_h$ and the $\tau_0$ parameter thus defines both globally and locally the smallest diffusion time scale of the problem. 

Nonetheless, all these differences are spurious once angular-resolution effects are taken into account.
More concretely, we verified that the luminosities (and flux densities) associated with our exact computation of the emissivities are reproduced by the RA formula. 
Specifically, we find for decaying DM in a NFW halo a $\sim$15\% difference between the corresponding computations of the flux density. 
In the case of annihilating DM this number is halved and for a point-mass scenario, the difference is zero.

In the middle and lower panels of fig. \ref{fig:exact} we consider $\tau_0$ parameters that are up to an order of magnitude smaller than the one considered in the upper panels. 
We also consider a completely different DM particle which is two orders of magnitude heavier and which primarily annihilates (decays) into $W^+W^-$ pairs.
Despite these huge differences on the model parameters, we observe that some of the characteristic features of the emissivity profiles are preserved.
For instance, the profiles shown in the middle panels still mimic the radial dependence that is predicted by the RA formula.
This certainly happens in a more modest $r$-range than in the upper panels but it is still appreciable and quantifiable: the RA flux-density predictions differ with the corresponding exact ones by 14\% (28\%) for a NFW profile in the case of DM annihilation (decay) and 41\%, 0.5\% for the Burkert and point-mass profiles respectively. 
In contrast, when compared to the RC formula we find that these yield flux densities that are four (Burkert) to eight (point mass) times larger than the exact computation.

In the lower panels instead, these departures are less dramatic.
Concretely, these amount to $\sim$~110\% for NFW (DM decay or annihilation), 95\% for Burkert and 150\% for point-mass profiles.
Also similar to what we found in our first benchmark case (uppermost panels), all profiles with cusps approach the RC result in the small-$r$ regime.
Note also that the RC and RA predictions of the middle lower panels are much closer to each other in all halo models than the ones obtained in the upper ones.
This is an manifestation that the transition between regimes is rather smooth.

Our results also confirm that the prescription that we provided earlier in order to define the different regimes of fig. \ref{fig:contours} is a reasonable one.
This can also be confirmed by observing that as the $\tau_0$ parameter is decreased, the vertical black lines move away from those associated with the RA computation (dashed blue) towards the RC (dotted green) ones.
We also observe that for the rightmost panels (PM case), the emissivity becomes less and less concentrated around $r=0$ as the parameter $\tau_0$ is decreased. 
We stress the fact that the previous remarks are largely independent of the DM particle mass and model.

Another rather interesting finding is that, independent of the DM model, the LMA (RB formula) for decaying DM in NFW halos performs very well in the transition regions (e.~g. middle and lower panels of the figure).
Indeed, the integrated flux-densities differ by only 3-7\% in the $\tau_0=4-15$~Gyr scenarios while the relative difference between the corresponding HFRs amounts to 6\%.
As already noticed in our discussion of fig. \ref{fig:halos} this curious fact is due to the suppression of the even harmonics in the series \eqref{eq:emissexact} (see relevant entries on table \ref{tab:hfactor}).

\section{Example. Draco limits on DM annihilation/decay}
\label{sec:draco}
In this section we propose a strategy to obtain reliable bounds on the annihilation cross section or decay rate of the DM by using the data (e.~g. noise level) of a given radio observation in a computationally efficient way.
There might also exist special situations in which a hypothetical detection could be interpreted as a discovery of the synchrotron emission from DM annihilation or decay.
However, their discussion is left for future work.

Our object of interest for this example is the Draco dwarf galaxy (R.~A. 17${}^{\rm h}~$20${}^{\rm m}$~12.4${}^{\rm s}$ and Dec. +57${}^{\rm \circ}$~54'~55'' \cite{1999ApJS..125..409C}), which is the first galaxy of its type that has been studied in the DM radio-detection context (e.~g. \cite{Tyler:2002ux,Colafrancesco:2006he}).
For definiteness, we assume that the galaxy has a half-light radius of $r_\star=$~320~pc. 
Its distance to the solar system is $R=$~75~kpc and we also assume that its mass distribution can be parametrized as a NFW halo with $4\rho_s=10^{-1.9}$~M${}_\odot/$pc${}^3$ ($\simeq0.48$~GeV/cm${}^3$) and $r_s=10^{3.46}$~pc ($\simeq2.9$~kpc $\simeq9\times r_\star$).
These parameters are consistent with \cite{Horigome:2020kyj}.

Given some raw observational data of Draco at a frequency $\nu_{\rm obs.}$, our strategy consists of modelling the DM signal in terms of only two parameters: the signal strength and the angular diameter.
The signal-shape parametrization will certainly also depend on the regime (A, B or C), the DM mass function (NFW, Einasto, etc.) and, of course, on whether the process of interest is annihilation or decay.

In particular, in our hypothetical Draco analysis the suggested templates would be then given by the regime-specific brightness profiles for NFW halos that are shown in tables \ref{tab:bhalodec} and \ref{tab:bhaloann} times signal-strength parameters $\mn_j^{\rm dec./ann.}$, where $j=$ A, B or C. 
For example, for annihilating DM in the regime C we would have to model the signal brightness as
\begin{equation}
I_{\nu_{\rm obs.},\, C}^{\rm ann.}(\theta)=\mn_C^{\rm ann.}\frac{32\pi^2\rho_s^2r_s^2}{r_h}\left[\sqrt{1-\frac{\theta^2}{\theta_h^2}}+\frac{\theta}{\theta_h}\left(\arctan\frac{\theta}{\sqrt{\theta_h^2-\theta^2}}-\frac{\pi}2\right)+\mathcal O(r_h/r_s)\right]\ .
\end{equation}
where $2\theta_h$ ($\simeq 2r_h/R$) is the angular diameter of the signal and the halo parameters $\rho_s$ and $r_s$ are given above.
Thus, the only parameters left on which the signal template can depend are $r_h$ or, equivalently $\theta_h$ and --in this particular example-- $\mn_C^{\rm ann.}$.
This is a tremendous simplification with respect to an otherwise cumbersome and \emph{redundant} multi-parameter scan of generic CRE propagation and particle DM models.  

Angular-diameter scans can be guided by educated guesses where $r_h$ ($R\theta_h$) is varied around a central value of $r_h^{\rm c}=r_\star$ ($=315$~pc for Draco).
If no statistically-significant excess is found in any of the scans and, say, $2\sigma$ limits on $\mn_j$ for each $j$, recasting these into synchrotron bounds on any DM model is a straightforward task.
Specifically, given set of the transport-model parameters ($B$, $\delta$, etc.) and the DM model that we want to test, we can start by identifying which regime $j=$~A, B or C that given set belongs to.
Then, we compute the model-specific spectral function $\mx_j(\nu_{\rm obs.})$. 
As a final step, the pre-factor $\mq$ in e.~g. \eqref{eq:brightmaster} (or \eqref{eq:emissmaster}) will be constrained by employing the following formula   
\begin{equation}
\mq_{2\sigma}^{\rm dec./ann. }=\frac{4\pi\,\bar{\mn}_j^{\rm dec./ann. }(\theta_h)}{X_j(\nu_{\rm obs.})}\ ,
\end{equation}
where $\bar{\mn}_j^{\rm dec./ann. }(\theta_h)$ is the value of the signal-strength parameter $\mn_j^{\rm dec./ann.}$ that has been excluded at  2$\sigma$ assuming a fixed value of $\theta_h$.
In particular, for annihilating self-conjugate DM we have that $\mq^{\rm ann.}=\langle\sigma v\rangle/(2\mchi^2)$ and the corresponding the annihilation rate limits would thus be given by 
\begin{equation}
\langle\sigma v\rangle_{2\sigma}=\frac{8\pi\mchi^2\,\bar{\mn}_j^{{\rm ann. }}(\theta_h)}{X_j(\nu_{\rm obs.})}\ .
\end{equation}

\subsection{Limit estimates}
Another approach that has been employed in, for instance, \cite{Leite:2016lsv,2020MNRAS.496.2663V} consists of estimating the limits on the synchrotron flux in terms of the noise level of the observation and the synthesized angular beam sizes. 
In this case, given a DM model and an appropriately chosen set of parameters the 2$\sigma$ limits on $\mq$ can be estimated by the following formula
\begin{eqnarray}
\bar{\mq}_{j\, 2\sigma}^{\rm dec./ann. } &=& 1.64\times\frac{4\pi}{\mh_\mr^j(r_h)\,X_j(\nu_{\rm obs.})}\frac{I_{\nu_{\rm obs.}}^{\rm r.m.s.}\times\Omega(r_h)}{\sqrt{\Omega({\rm HFR})/\Omega_{\rm beam}}}\nonumber\\
\label{eq:bounds}
{} &\simeq&20.6\times
\frac{I_{\nu_{\rm obs.}}^{\rm r.m.s.}\Omega_{\rm beam}}{\mh_\mr^j(r_h)\,X_j(\nu_{\rm obs.})}\left(\frac{r_h}{\rm HFR}\right)\sqrt{\frac{\Omega(r_h)}{\Omega_{\rm beam}}}\ ,
\end{eqnarray}
where $I^{\rm r.m.s.}_{\nu_{\rm obs.}}$ is the root-mean-square brightness noise of the observation, $\Omega_{\rm beam}$ is the beam size and $\Omega(r_h)\simeq\pi(r_h/R)^2$ is the angular size of the emitting region and we estimate the number of beams as $\Omega({\rm HFR})/\Omega_{\rm beam}$ where $\Omega({\rm HFR})\simeq\pi({\rm HFR}/R)^2$. 
The halo factors $\mh_\mr$ and the HFRs are given in the appendix as well as the derivation of the formula \eqref{eq:bounds}.

Using the parameters for Draco halo model we can obtain these as functions of the angular radius $\theta_h$.
The results for our three approximations are listed in \ref{tab:dracohs}. 
In particular, the RA entry for DM annihilation agrees (within uncertainties) with the $J$-factor reported in \cite{Horigome:2020kyj}.

\begin{table}[h!]
\centering{
\begin{tabular}{|c||c|c|}
\hline
\rule{0pt}{15pt}
\bf Regime & \boldmath\bf Decay ($\times 10^{17}$~GeV/cm${}^2$) & \boldmath  \bf Annihilation ($\times 10^{18}$~GeV${}^2$/cm${}^5$)\\[5pt]
\hline
\hline
\rule{0pt}{15pt}
A & $20.6\left(\frac{\theta_h}{0.5^\circ}\right)^2\left[1-0.30\left(\frac{\theta_h}{0.5^\circ}\right)+\ldots\right]$  & $8.74\left(\frac{\theta_h}{0.5^\circ}\right)\left[1-0.45\left(\frac{\theta_h}{0.5^\circ}\right)+\ldots\right]$ \\[5pt]
\rule{0pt}{15pt}
B  & $16.7\left(\frac{\theta_h}{0.5^\circ}\right)^2\left[1-0.23\left(\frac{\theta_h}{0.5^\circ}\right)+\ldots\right]$ &  $3.28\left(\frac{\theta_h}{0.5^\circ}\right)\left[1-0.31\left(\frac{\theta_h}{0.5^\circ}\right)+\ldots\right]$\\[5pt]
\rule{0pt}{15pt}
C  & $1.71\left(\frac{\theta_h}{0.5^\circ}\right)^2\left[1-0.24\left(\frac{\theta_h}{0.5^\circ}\right)+\ldots\right]$ & $0.97\left(\frac{\theta_h}{0.5^\circ}\right)\left[1-0.34\left(\frac{\theta_h}{0.5^\circ}\right)+\ldots\right]$ \\[5pt]
\hline
\end{tabular}
}
\caption{\label{tab:dracohs} Flux-density halo factors for Draco using the NFW parametrization with $\rho_s=0.12$~GeV/cm${}^3$ and $r_s=2.9$~kpc.}
\end{table}

\begin{figure}
\includegraphics[width=\linewidth]{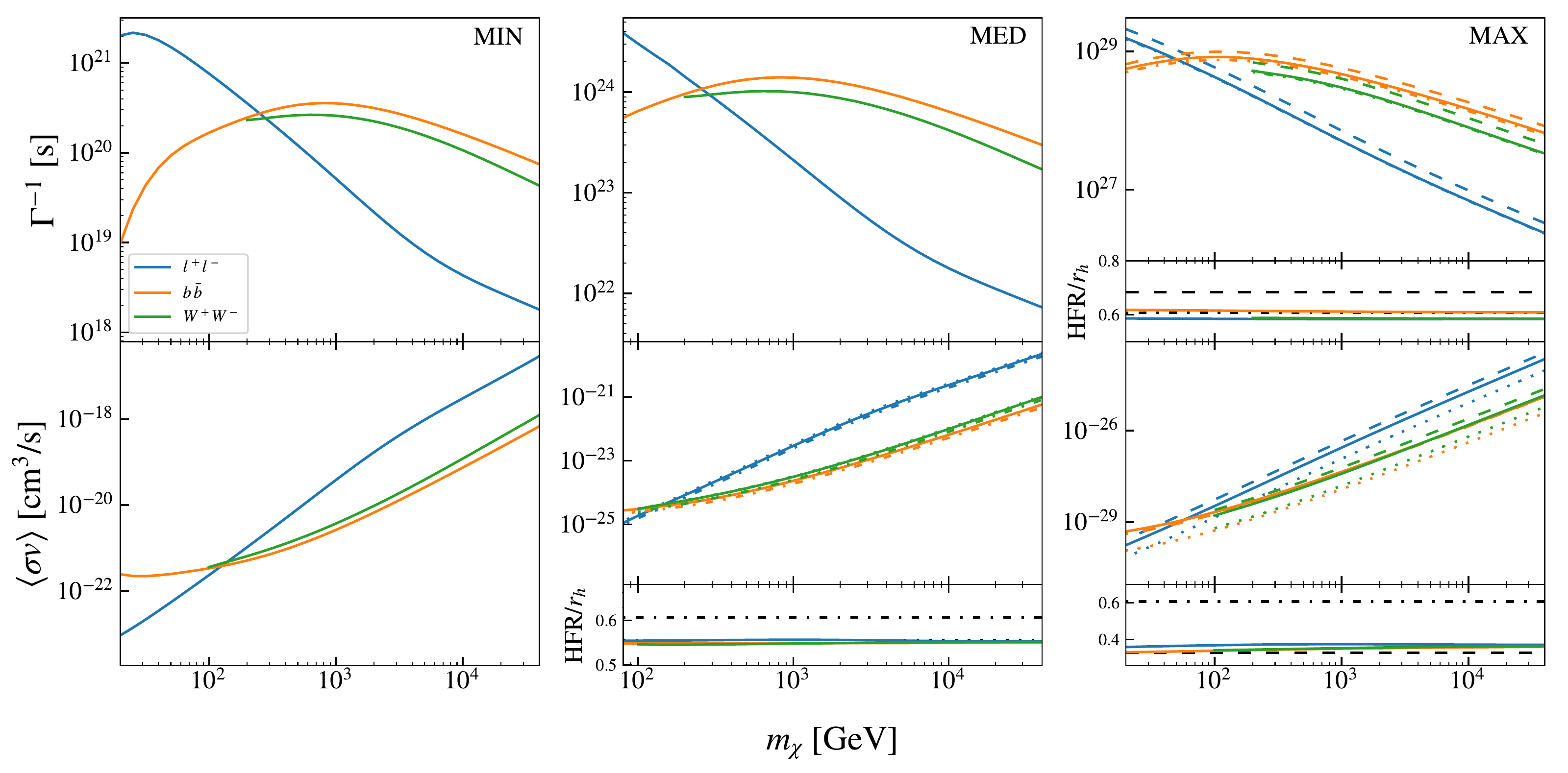}
\caption{\label{fig:dracofictlimits} Dummy Draco limits on the decay (annihilation) rates for annihilating and decaying DM. 
Solid lines show the exact computations.
Different colors indicate different assumptions on the leading hard process in the annihilation or decay. 
Three models are considered: $l^+l^-$ (blue), $b\bar b$ (orange) and $W^+W-$ (green).
For the MED and MAX scenarios we also included the exact values for the HFRs.
Dashed lines indicate whether the RA approximation was employed. 
Analogously dot-dashed and dotted lines are respectively associated with RB and RC. 
 }
\end{figure}

In order to visualize how such limit estimates look like, we assume fictitious 150~MHz observations with $I_{150\,\rm MHz}^{\rm r.m.s.}=100\,\mu$Jy/beam where the beam size is assumed to be of 20''. 
Fig. \ref{fig:dracofictlimits} shows the estimated Draco limits on several DM models that can be extracted from such observations.  
For concreteness, we consider three CRE-transport parameter sets MIN, MED and MAX.
These are outlined in the table below
\medskip

\begin{table}
\centerline{
\begin{tabular}{|c|c|c|c||c|c|c|}
\hline
\rule{0pt}{10pt}
\bf Model & \boldmath$D_0$ & \boldmath$r_h$ ($\theta_h$)& \boldmath$\tau_0$ & \boldmath$B$ & \boldmath$\delta$ & \bf Regime \\[1pt]
\hline
\hline
\rule{0pt}{10pt}
MIN & 3~$\times10^{29}$~cm${}^2$/s & 160~pc (0.12${}^\circ$)& 25~kyr & 0.1~$\mu$G & 0.3 & C\\[5pt]
\rule{0pt}{5pt}
MED & 3~$\times10^{27}$~cm${}^2$/s & 640~pc (0.49${}^\circ$) & 40~Myr & 0.1~$\mu$G & 0.3 & B/C\\[5pt]
\rule{0pt}{5pt}
MAX & 3~$\times10^{25}$~cm${}^2$/s & 1~kpc (0.76${}^\circ$) & 10~Gyr & 10~$\mu$G & 0.3 & A/B \\[5pt]
\hline
\end{tabular}
}
\caption{Adopted transport-equation parameters in our MIN, MED and MAX scenarios for Draco.}
\end{table}
\bigskip

A few comments on the resulting limits are in order.
First, models whose primary annihilation/decay process are the final states $W^+W^-$ and $b\bar b$ have a characteristic mass  around 1~TeV at which limits on the decay rate become maximal. 
For leptophilic DM we see a similar behaviour but at a much smaller mass $\mathcal O$(10GeV). 
Physically, what happens in those models ($W^+W^-$ and $b\bar b$ final states) is that since most of the electrons that are dumped by the annihilation/decay are much less energetic than the DM mass, the characteristic frequency at which the bulk of the electrons can maximally emit is much smaller than the corresponding frequency in e.~g. the leptophilic scenario.

The RC formula is appropriate for the MIN and MED scenarios. 
Though, in the case of DM annihilation, $\mathcal O(10\%)$ differences are present. 
In the MAX case, we observe that in the DM-decay scenario, the RB formula performs much better than the RA one in (1) reproducing the exact limits and (2) the exact HFRs.
The latter condition is a consequence that the emissivity profiles have similar shapes.
This is an accidental property that has been noted in the previous sections.
The reason is that the contribution of even modes --particularly the $n=2$ mode-- in the series \eqref{eq:emissexact} is rather suppressed.

Due to the huge uncertainties on the transport-model parameters on dwarf galaxies, the non-observation of diffuse signals in those galaxies could lead to limits on e.~g. the annihilation cross section of WIMPs that are so strong that a large chunk in the parameter-space of thermally-produced ($s$-wave) DM models would be ruled out.
However, our estimates are also compatible with the hypothesis that synchrotron signals from e.~g.  WIMP annihilation are much too weak to be probed by existing radio telescopes by simply adopting different assumptions on the parameters.
This is of course quite unsatisfactory and it calls for an intensification of both theoretical and experimental effort in reducing these uncertainties.

\section{Conclusions}
We developed new methods for the computation of the rather uncertain synchrotron fluxes from dwarf galaxies. 
Focus was placed on the situation where the leading process giving rise to these signals is the annihilation or decay of dark-matter particles.
By identifying three regimes where approximations are possible, our main result is the derivation of regime-dependent formulas for the synchrotron signals where frequency and DM-model independent halo functions can be defined.
These functions only depend on the $r_h$ parameter and the DM distribution.
And as a consequence, all the frequency and DM-model dependence of the formula is captured in a spectral term which we also compute for generic scenarios.
In particular, we described a numerical method in order to carry out these computations very efficiently.
Further, we proposed a strategy for the search of synchrotron signals in dwarf galaxies based on two-parameter signal templates.
This is a strong simplification of the otherwise very complicated multi-parameter analysis.

We conducted several tests in order to assess the validity of our regime-A, B and C approximations.
For this task, we took advantage of the fact that the computations of the exact solutions are straightforward in our setup.
We also carried out several consistency checks and discussed them. 
In particular, we corrected a mistake in the alternative method-of-images expression for the Green's function that has been employed in the literature for quite some time.

Due to the increasing observational activity in these directions and in light of e.~g. the Square Kilometre Array, we identify several challenges for future theoretical research.
The most urgent is, certainly, advancing our theoretical knowledge about the evolution of magnetic fields in galaxies where star formation is no longer active.
From the observational side, new and deeper observations are essential as well as the sophistication of the strategies.
Concerning the CRE propagation modelling and light DM candidates, the effect of re-acceleration and halo evolution (departure from the stationary condition) becomes important and should be addressed in future studies. 
\bigskip

{\bf Acknowledgements.} The author would like to thank Torsten Bringmann and Marcus Br\"uggen for carefully reading this manuscript as well as for their very valuable comments, Rainer Beck for pointing out an argument motivating the presence of sizeable magnetic fields in dwarf galaxies in the present time and to Stefano Profumo and Piero Ullio for their readiness to cross check our results. 
Thanks also to Volker Heesen, G\"unter Sigl, Alejandro Ibarra and Martin Beneke for their feedback and the anonymous referee for their valuable comments and for verifying (numerically) several claims made in this paper.
In particular, the correctness of \eqref{eq:images}. 
A substantial part of this work was supported by the DFG Collaborative Research Centre ``Neutrinos and Dark Matter in Astro- and Particle Physics'' (SFB 1258).

\appendix
\section{Complementary analytic derivations}
\label{app:analytics}
In this appendix we sketch the derivation of our Fourier expansion of the transport-equation Green's function \eqref{eq:fourier} and its effective expressions in the limits of no diffusion and rapid diffusion.
We start by writing down differential equation for the Green's function $G_{r_h}^{\rm 1D}(\Delta\lambda^2,r,r')$ (henceforth $\tilde G(\Delta\lambda^2,r,r')$):
\begin{equation}
\label{eq:diffeqgreen}
\frac{\partial}{\partial\lambda^2}\tilde G(\Delta\lambda^2,r,r')-\frac{\partial^2}{\partial r^2}\tilde G(\Delta\lambda^2,r,r')=\delta(\Delta\lambda^2)\delta(r-r')\quad , \quad \tilde G(\Delta\lambda^2,r_h,r')=0\ .
\end{equation}

Here we will use for simplicity the symmetry $\tilde G(\Delta\lambda,r,r')=\tilde G(\Delta\lambda,r',r)$.
This is justified by noting that the method-of-images expansion of the Green's function \eqref{eq:images} also has this symmetry. 
We are certainly implicitly invoking theorems of existence and uniqueness of the solutions of the heat equation.
As a consequence of the aforementioned property and the boundary conditions on $r=0$ ($r'=0$) and $r=r_h$  ($r'=r_h$) we may, without loss of generality, expand $\tilde G$ as
\begin{equation}
\label{eq:ansatz}
\tilde G(\Delta\lambda^2,r,r')=\sum_{m,n=0}^\infty\Theta(\Delta\lambda^2) f_{nm}(\Delta\lambda^2)\sin\frac{m\pi r}{r_h}\sin\frac{n\pi r'}{r_h}\ ,
\end{equation}
where, by construction, the coefficients $f_{mn}(\Delta\lambda^2)$ satisfy
\begin{equation}
\frac{\dif}{\dif\lambda^2}f_{mn}(\Delta\lambda^2)+\frac{m^2\pi^2}{r_h^2}f_{mn}(\Delta\lambda^2)=0\ .
\end{equation}
Thus, $f_{mn}(\Delta\lambda^2)=c_{mn}\e^{-\frac{m^2\pi^2\Delta\lambda^2}{r_h^2}}$. 
The coefficients $c_{mn}$ are determined by using the $r\leftrightarrow r'$ symmetry again and the right-hand side of \eqref{eq:diffeqgreen}.
Namely, the former condition implies that $c_{mn}=C\times \delta_{mn}$ ($\delta_{mn}$ is a Kroenecker delta and $C$ a constant), while the latter one yields $C=2/r_h$. 
Concretely, by inserting our ansatz \eqref{eq:ansatz} into \eqref{eq:diffeqgreen} and using the identity
\begin{equation}
\label{eq:identity}
\sum_{n=1}^\infty\sin(nx)\sin(ny)=\frac{\pi}2\delta(x-y)\ .
\end{equation}

Thus,
\begin{equation}
 \delta(\Delta\lambda^2)\times C\times\sum_{n=1}^\infty\sin\frac{n\pi r}{r_h}\sin\frac{n\pi r'}{r_h}= \frac{r_h\,C}2\delta(\Delta\lambda^2)\delta(r-r')=\delta(\Delta\lambda^2)\delta(r-r')\ ,
\end{equation}
where the last equation is of course satisfied provided that $C=2/r_h$. 

\subsection{Regime-A and -C limits}
Using the identity \eqref{eq:identity} we can almost immediately obtain the RA formula \eqref{eq:regimeA} by explicitly evaluating $\Delta\lambda=0$ in \eqref{eq:fourier}.
However, notice that although the limit reproduces the RA result, the leading-power correction of the RA formula that is (na\"{\i}vely) computed by Taylor expanding the exponential terms $\e^{-n^2\Delta\eta}$ in \eqref{eq:fourier}, yields a divergent result and a more sophisticated expansion is required.

In the opposite limit where the variable $\lambda$ is (typically) much larger than $r_h$, or $\eta$ much larger than unity, we may use the following expansion
\begin{equation}
\label{eq:identitytwo}
\Theta(z)\e^{-\alpha z}\to\frac1{\alpha}\delta(z)-\frac1{\alpha^2}\delta'(z)+\frac1{\alpha^3}\delta''(z)+\ldots=\delta(\alpha z)-\delta'(\alpha z)+\delta''(\alpha z)+\ldots\ ,
\end{equation}
which is valid for large values of $\alpha>0$. 
Its validity can be proven by evaluating the integral of the multiplication of a Taylor-expanded test function $f(z)$ together with the left-hand side of the equation above over any interval including $z=0$, e.~g. $[-Z_-,Z_+]$:
\begin{eqnarray}
\int_{-Z_-}^{Z_+}\dif z\,\Theta(z)\e^{-\alpha z}f(z)&=&\int_0^{\alpha Z_+}\frac{\dif u}{\alpha}\e^{-u}\left(f(0)+\frac{u}{\alpha}f'(0)+\frac{u^2}{2\alpha^2}f''(0)+\ldots\right)\\
{} &\approx&\int_0^{\infty}\frac{\dif u}{\alpha}\e^{-u}\left(f(0)+\frac{u}{\alpha}f'(0)+\frac{u^2}{2\alpha^2}f''(0)+\ldots\right) \\
{}&=&\frac1{\alpha}\left(f(0)+ \frac1{\alpha}f'(0)+ \frac1{\alpha^2}f''(0)+\ldots\right)\ .
\end{eqnarray} 

In the following, we will use the identity \eqref{eq:identitytwo} in order to reproduce the regime-C formula and its leading-power correction.
Namely,
\begin{eqnarray}
\Theta(\Delta\eta)\e^{-n^2\Delta\eta} &=&\delta(n^2\Delta\eta)-\delta'(n^2\Delta\eta)+\ldots=\frac1{n^2}\delta(\Delta\eta)-\frac1{n^4}\delta'(\Delta\eta)+\ldots\\
{} &=&\frac{r_h^2}{n^2\pi^2}\delta(\Delta\lambda^2)-\frac{r_h^4}{n^4\pi^4}\delta'(\Delta\lambda^2)+\ldots\\
{} &=&\frac{r_h^2\,b(E)}{n^2\pi^2D(E)}\delta(\Delta E)-\frac{r_h^4\,b^2(E)}{n^4\pi^4D^2(E)}\delta'(\Delta E)+\ldots\ ,
\label{eq:powexp}
\end{eqnarray}
where in the last step we used \eqref{eq:syrovatskii}. 
In particular, by using the last equality (and $n=1$) the reader can verify that the expression \eqref{eq:specc} is indeed obtained when \eqref{eq:specb} is evaluated at very large $\eta$.
As a final remark, note that in contrast with the small-$\lambda$ limit, all power corrections can be easily computed by using the sub-leading terms of the expansion \eqref{eq:powexp} with e.~g. $\tau_0=r_h^2/D_0$ as expansion parameter.

\section{Brightness and flux-density halo factors}
\label{app:brighthalos}

\begin{table}[t!]
\centering{
\begin{tabular}{|c|c|c|c|}
\hline
\rule{0pt}{10pt}
\bf Halo model & \bf A & \bf\boldmath B${}^{(1)}$ &\bf C\\[2pt]
\hline
\hline
\rule{0pt}{12pt}
PM & $(M_h/R^2)\times$~PSF$(\theta)$ & $\frac{M_h}{r_h^2}f_B(t)$ &  $\frac{\pi\,M_h}{2 r_h^2}\left[\ln\frac{1+\sqrt{1-t^2}}t-\sqrt{1-t^2}\right]$ 
\\[5pt]
\hline
\rule{0pt}{12pt}
C  & $2\rho_hr_h\sqrt{1-t^2}$ & $\frac{4r_h\rho_h}{\pi}f_B(t)$ & $\frac{2\pi^2\rho_hr_h}9(1-t^2)^{3/2}$ \\[5pt]
\hline
\rule{0pt}{20pt}
NFW &\begin{tabular}{c} 
$8\rho_s r_s\left(\ln\frac{1+\sqrt{1-t^2}}t\right.$\\
$\left.-2\frac{r_h}{r_s}\sqrt{1-t^2}+\ldots\right)$ 
\end{tabular}
& $2\, h_{\rm NFW}^{\rm dec.}f_B(t)$ & \parbox[c]{2.2in}{\centering$2\pi^2\rho_sr_s\left[\sqrt{1-t^2}-t^2\tanh^{-1}\sqrt{1-t^2}\,+\right.$ $\left.-\frac{8\,r_h}{9\,r_s}(1-t^2)^{3/2}+\ldots\right]$ .
} \\[12pt]
\hline
\rule{0pt}{25pt}
Bkrt & \parbox[c]{1.7in}{\centering$2\tilde\rho_sr_h\sqrt{1-t^2}\left[1-\frac{r_h}{2\tilde r_s}\Big{(}1+\right.$	
$\left.\left.\ +\ t^2\frac{\tanh^{-1}\sqrt{1-t^2}}{\sqrt{1-t^2}}\right)+\ldots\right]$} & $2\,h_{\rm Bkrt}^{\rm dec.}f_B(t)$ & 
\begin{tabular}{c}
$\frac{2\,\pi^2}9\tilde\rho_s r_h\sqrt{1-t^2}\left[1-t^2\right.$\\
$\left.-\frac{9\,r_h}{32\tilde r_s}\left(2-t^2-t^4\frac{\tanh^{-1}\sqrt{1-t^2}}{\sqrt{1-t^2}}\right)+\ldots\right]$
\end{tabular}
\\[15pt]
\hline
\end{tabular}}
\caption{\label{tab:bhalodec} Brightness halo functions $\hat H_\rho(\theta)$ for DM decay.}
\end{table} 

In section \ref{sec:universal} we provided the emissivity halo functions for several benchmark parametrizations of the dSph halos.  
In this appendix we turn our attention to their associated brightness and flux densities.
Since absorption effects are neglected, this is straightforward.
Namely, the generalization of \eqref{eq:emissmaster} reads
\begin{eqnarray}
\label{eq:brightmaster}
I_\nu(\theta) &=&\frac{\mq}{4\pi}\times \hat H_\mr(\theta)\times \mx(\nu)\ ,\\
\label{eq:fluxdmaster}
S_\nu &=&\frac{\mq}{4\pi}\times \mh_\mr\times \mx(\nu)\ ,\\
\end{eqnarray}
where, in analogy with \eqref{eq:emissmaster}, $\hat H_\mr$ and $\mh_\mr$ are the regime-dependent brightness and flux-density halo functions of the synchrotron signal.
The brightness ($\hat H_\mr$) and flux-density ($\mh_\mr$) halo functions are obtained from $H_\mr$ as follows
\begin{eqnarray}
\hat H_\mr(\theta) &=& \int_{-\sqrt{r_h^2-R^2\sin^2\theta}}^{\sqrt{r_h^2-R^2\sin^2\theta}}\!\dif l\,H_\mr\left(\sqrt{l^2+R^2\sin^2\theta}\right)=\int_{R\sin\theta}^{r_h}\frac{2\, rH_\mr(r)\dif r}{\sqrt{r^2-R^2\sin^2\theta}}\ ,\\
\mh_\mr &=& 2\pi\int_{0}^{\arcsin\frac{r_h}R}\!\dif \theta\,\sin\theta\,\hat H_\mr\left(\theta\right)\ .
\end{eqnarray}
or, more appropriately in terms of the variable $t=R\sin\theta/r_h\approx\theta/\theta_h$.
\begin{eqnarray}
\label{eq:bhalot}
\hat H_\mr(t) &=& \int_{-r_h\sqrt{1-t^2}}^{r_h\sqrt{1-t^2}}\!\dif l\,H_\mr\left(\sqrt{l^2+r_h^2t^2}\right)=\int_{r_ht}^{r_h}\frac{2\, rH_\mr(r)\dif r}{\sqrt{r^2-r_h^2t^2}}\ ,\\
\label{eq:fdhalo}
\mh_\mr &=& \frac{2\pi r_h^2}{R^2}\int_0^1\!\frac{\dif t\,t}{\sqrt{1-\frac{r_h^2 t^2}{R^2}}}\,\,\hat H_\mr(t)=\frac{2\pi r_h^2}{R^2}\int_0^1\!\dif t\,t\,\,\hat H_\mr(t)+\mathcal O(r_h^4/R^4)\ .
\end{eqnarray}
Note that in our nomenclature, the flux density is integrated over the whole galaxy. 
In particular, in the regime B 
\[\hat H_\mr^B(\theta)\equiv 2\,h_\mr f_B(t)\ ,\]
where
\[f_B(t)=\int_t^1\frac{\sin\pi \tau\,\dif \tau}{\sqrt{\tau^2-t^2}}\ .\]

\begin{table}[t!]
\centering{
\begin{tabular}{|c|c|c|c|}
\hline
\rule{0pt}{10pt}
\bf Halo model & \bf  A & \bf \boldmath B${}^{(1)}$  &\bf C\\[2pt]
\hline
\hline
\rule{0pt}{12pt}
PM & $(K_{\rm pm}/R^2)\times$~PSF$(\theta)$ & $\frac{K_{\rm pm}}{r_h^2}f_B(t)$ &  $\frac{\pi\,K_{\rm pm}}{2 r_h^2}\left[\ln\frac{1+\sqrt{1-t^2}}t-\sqrt{1-t^2}\right]$ 
\\[5pt]
\hline
\rule{0pt}{12pt}
C  & $2\rho_h^2r_h\sqrt{1-t^2}$ & $\frac{4r_h\rho_h^2}{\pi}f_B(t)$ & $\frac{2\pi^2\rho_h^2r_h}9(1-t^2)^{3/2}$ \\[5pt]
\hline
\rule{0pt}{25pt}
NFW &
\parbox[c]{1.7in}{\centering
 $\frac{32\rho_s^2 r_s^2}{r_h}\left(\frac1t\arctan\frac{\sqrt{1-t^2}}t\right.$
 $\left.-4\frac{r_h}{r_s}\tanh^{-1}\sqrt{1-t^2}+\ldots\right)$} & $2\,h_{\rm NFW}^{\rm ann.}f_B(t)$ & \parbox[c]{2.4in}{\centering$\frac{32\pi^2\rho_s^2 r_s^2}{r_h}\left[\sqrt{1-t^2}+t\left(\arctan\frac{t}{\sqrt{1-t^2}}-\frac{\pi}2\right)\right.$ 
$\left.-\frac{r_h}{r_s}\left(\sqrt{1-t^2}-t^2\tanh^{-1}\sqrt{1-t^2}\right)+\ldots\right]$ .} \\[15pt]
\hline
\rule{0pt}{25pt}
Bkrt & \parbox[c]{1.7in}{\centering$2\tilde\rho_s^2r_h\sqrt{1-t^2}\left[1-\frac{r_h}{\tilde r_s}\Big{(}1+\right.$	
$\left.\left.\ +\ t^2\frac{\tanh^{-1}\sqrt{1-t^2}}{\sqrt{1-t^2}}\right)+\ldots\right]$} & $2\,h_{\rm Bkrt}^{\rm ann.}f_B(t)$ & \begin{tabular}{c}
$\frac{2\pi^2}9\tilde\rho_s^2 r_h\sqrt{1-t^2}\left[1-t^2\right.$\\
$\left.-\frac{9\,r_h}{16\tilde r_s}\left(2-t^2-t^4\frac{\tanh^{-1}\sqrt{1-t^2}}{\sqrt{1-t^2}}\right)+\ldots\right]$
\end{tabular}
\\[15pt]
\hline
\end{tabular}}
\caption{\label{tab:bhaloann} Brightness halo functions $\hat H_{\rho^2}(\theta)$ for DM annihilation.}
\end{table} 

The integral \eqref{eq:fdhalo} in this case is computable (in the $r_h\ll R$ limit) and yields
\[\mh_B=4\,h_\mr \frac{r_h^2}{R^2}\]

\begin{figure}
\includegraphics[width=\linewidth]{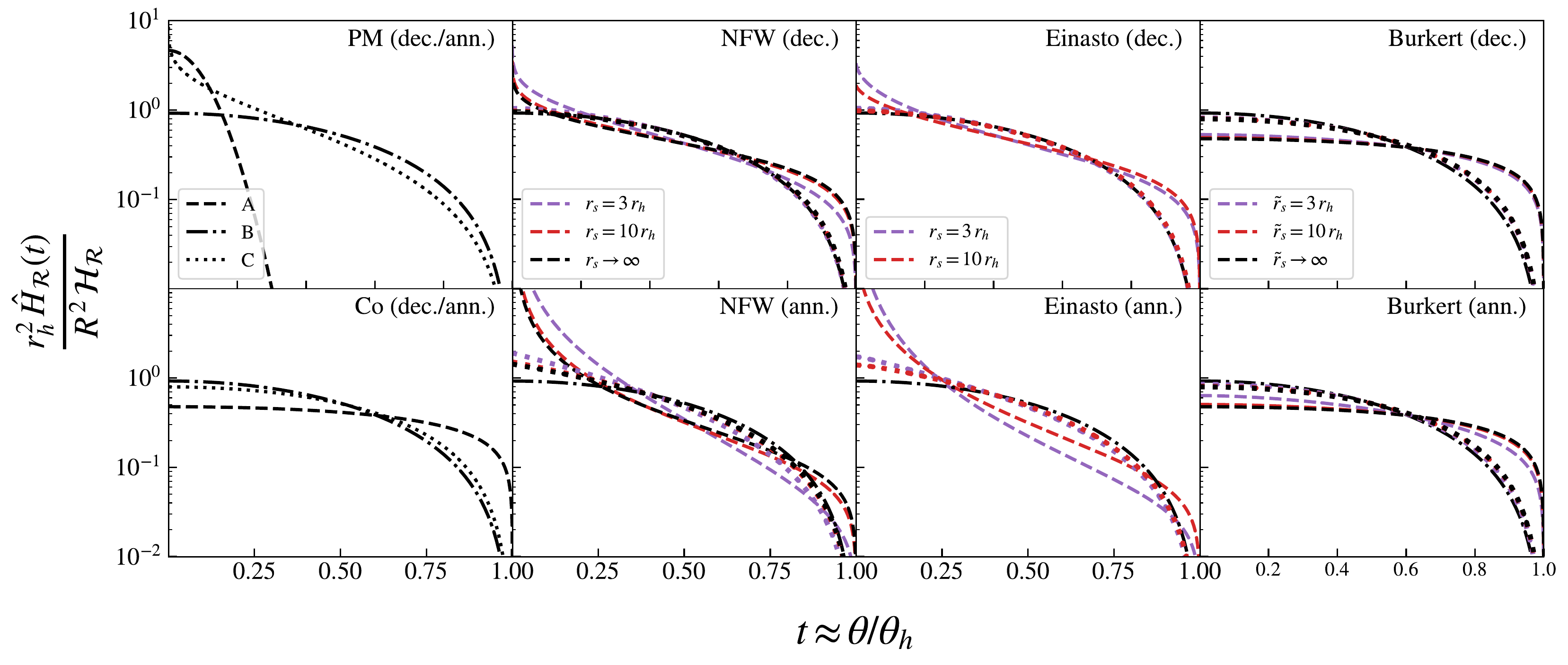}
\caption{\label{fig:bhalos} (Normalized) brightness halo functions for the several DM-mass profiles shown in fig. \ref{fig:halos}. In the NFW case and for annihilating DM, we included $\mathcal O(r_h^2/r_s^2)$ terms so that the positivity of the brightness is not jeopardized by the Taylor expansion.}
\end{figure}

Tables \ref{tab:bhalodec}-\ref{tab:bhaloann} are the analogues of \ref{tab:halodec}-\ref{tab:haloann} for the brightness, i.~e. brightness halo factors --as functions of $t$-- for several benchmark parametrizations of the dSph halos.
These are computed using \eqref{eq:bhalot} and the entries of table \ref{tab:halodec} and \ref{tab:haloann}.
Likewise, we show in the table \ref{tab:fdhalo} the flux-density factors (obtained using \eqref{eq:fdhalo}) and the corresponding half-flux radii for the aforementioned halo models.
Fig. \ref{fig:bhalos} is the analogue of \ref{fig:halos}. 
Namely, the entries of tables \ref{tab:bhalodec}-\ref{tab:bhaloann} are plotted as functions of the variable $t$.
Note that we are also including the Einasto model, which brightness factors are obtained numerically.

\begin{sidewaystable}
\begin{tabular}{|c|c|c|c|c|c|c|}
\hline
\rule{0pt}{10pt}
Decay & \multicolumn{2}{|c|}{\bf  A} &  \multicolumn{2}{|c|}{\bf \boldmath B${}^{(1)}$}  &\multicolumn{2}{|c|}{\bf  C}\\[2pt]
\hline
\rule{0pt}{10pt}
\bf Model & $\mh_\rho$ & HFR &  $\mh_\rho$ & HFR &$\mh_\rho$ & HFR \\[2pt]
\hline
\hline
\rule{0pt}{12pt}
PM & $\frac{M_h}{R^2}$ & 0 & $\frac{2M_h}{R^2}$ & $0.607\,r_h$ & $\frac{\pi^2M_h}{6R^2}$ & $\frac{r_h}2$
\\[5pt]
\hline
\rule{0pt}{12pt}
C  & $\frac{4\pi r_h^3\rho_h}{3R^2}$ & $\frac{r_h}{2^{1/3}}$ & $\frac{8\rho_h r_h^3}{\pi R^2}$ & $0.607\,r_h$ & $\frac{4\pi^3 r_h^3\rho_h}{45R^2}$ & $0.64\, r_h$ \\[5pt]
\hline
\rule{0pt}{20pt}
NFW &
$8\pi\rho_s r_s\frac{r_h^2}{R^2}\left(1-\frac43\frac{r_h}{r_s}+\ldots\right)$ & $\left(\frac1{\sqrt{2}}-\frac{\sqrt{2}-1}6\frac{r_h}{r_s}+\ldots\right)r_h$ & $4h_{\rm NFW}^{\rm dec.}\frac{r_h^2}{R^2}$ & $0.607\,r_h$ & $\frac{2\pi^3\rho_s r_sr_h^2}{3R^2}\left(1-\frac{16}{15}\frac{r_h}{r_s}+\ldots\right)$ & $0.614\,r_h\left(1-0.051\frac{r_h}{r_s}+\ldots\right)$ \\[12pt]
\hline
\rule{0pt}{25pt}
Bkrt & $\frac{4\pi r_h^3\tilde\rho_s}{3R^2}\left(1-\frac34\frac{r_h}{\tilde r_s}+\ldots\right)$ & $\frac{r_h}{2^{1/3}}\left(1-\frac{(2^{1/3}-1)r_h}{4\times2^{2/3}\tilde r_s}+\ldots\right)$ & $4h_{\rm Bkrt}^{\rm dec.}\frac{r_h^2}{R^2}$ & $0.607\,r_h$ & $\frac{4\pi^3 r_h^3\tilde\rho_s}{45R^2}\left(1-\frac58\frac{r_h}{\tilde r_s}+\ldots\right)$ & $0.643\, r_h\left(1-0.021\frac{r_h}{r_s}\right)$  \\[15pt]
\hline
\end{tabular}
\vspace{50pt}

\begin{tabular}{|c|c|c|c|c|c|c|}
\hline
\rule{0pt}{10pt}
Annihilation & \multicolumn{2}{|c|}{\bf A} &  \multicolumn{2}{|c|}{\bf \boldmath B${}^{(1)}$}  &\multicolumn{2}{|c|}{\bf C}\\[2pt]
\hline
\rule{0pt}{10pt}
\bf Model & $\mh_{\rho^2}$ & HFR &  $\mh_{\rho^2}$ & HFR &$\mh_{\rho^2}$ &  HFR \\[2pt]
\hline
\hline
\rule{0pt}{12pt}
PM & $\frac{K_{\rm pm}}{R^2}$ & 0 & $\frac{2K_{\rm pm}}{R^2}$ & $0.607\,r_h$ &  $\frac{\pi^2K_{\rm pm}}{6R^2}$ & $\frac{r_h}2$
\\[5pt]
\hline
\rule{0pt}{12pt}
C  & $\frac{4\pi r_h^3\rho_h^2}{3R^2}$ & $\frac{r_h}{2^{1/3}}$ & $\frac{8\rho_h^2 r_h^3}{\pi R^2}$ & $0.607\,r_h$ & $\frac{4\pi^3 r_h^3\rho_h^2}{45R^2}$ & $0.643\, r_h$ \\[5pt]
\hline
\rule{0pt}{20pt}
NFW &
$\frac{64\pi\rho_s^2 r_s^2r_h}{R^2}\left(1-2\frac{r_h}{r_s}+\ldots\right)$ & $\frac{r_h}2\left(1-\frac{r_h}{r_s}+\ldots\right)$
& $4h_{\rm NFW}^{\rm ann.}\frac{r_h^2}{R^2}$  & $0.607\,r_h$ & $\frac{64\pi^3\rho_s^2 r_s^2r_h}{9R^2}\left(1-\frac32\frac{r_h}{r_s}+\ldots\right)$ & $0.572\left(1-0.12\frac{r_h}{r_s}+\ldots\right)$ \\[12pt]
\hline
\rule{0pt}{25pt}
Bkrt & $\frac{4\pi r_h^3\tilde\rho_s^2}{3R^2}\left(1-\frac32\frac{r_h}{\tilde r_s}+\ldots\right)$ & $\frac{r_h}{2^{1/3}}\left(1-\frac{(2^{1/3}-1)r_h}{4\times2^{2/3}\tilde r_s}+\ldots\right)$  & $4h_{\rm Bkrt}^{\rm ann.}\frac{r_h^2}{R^2}$  & $0.607\,r_h$ & $\frac{4\pi^3 r_h^3\tilde\rho_s^2}{45R^2}\left(1-\frac54\frac{r_h}{\tilde r_s}+\ldots\right)$ & $0.643\, r_h\left(1-0.042\frac{r_h}{r_s}+\ldots\right)$\\[15pt]
\hline
\end{tabular}
\caption{\label{tab:fdhalo} Flux density halo factors and their corresponding half-flux radii for DM decay and annihilation.}
\end{sidewaystable} 

\section{Generic particle DM models}
\label{app:spectra}
\begin{figure}[h!]
\includegraphics[width=\linewidth]{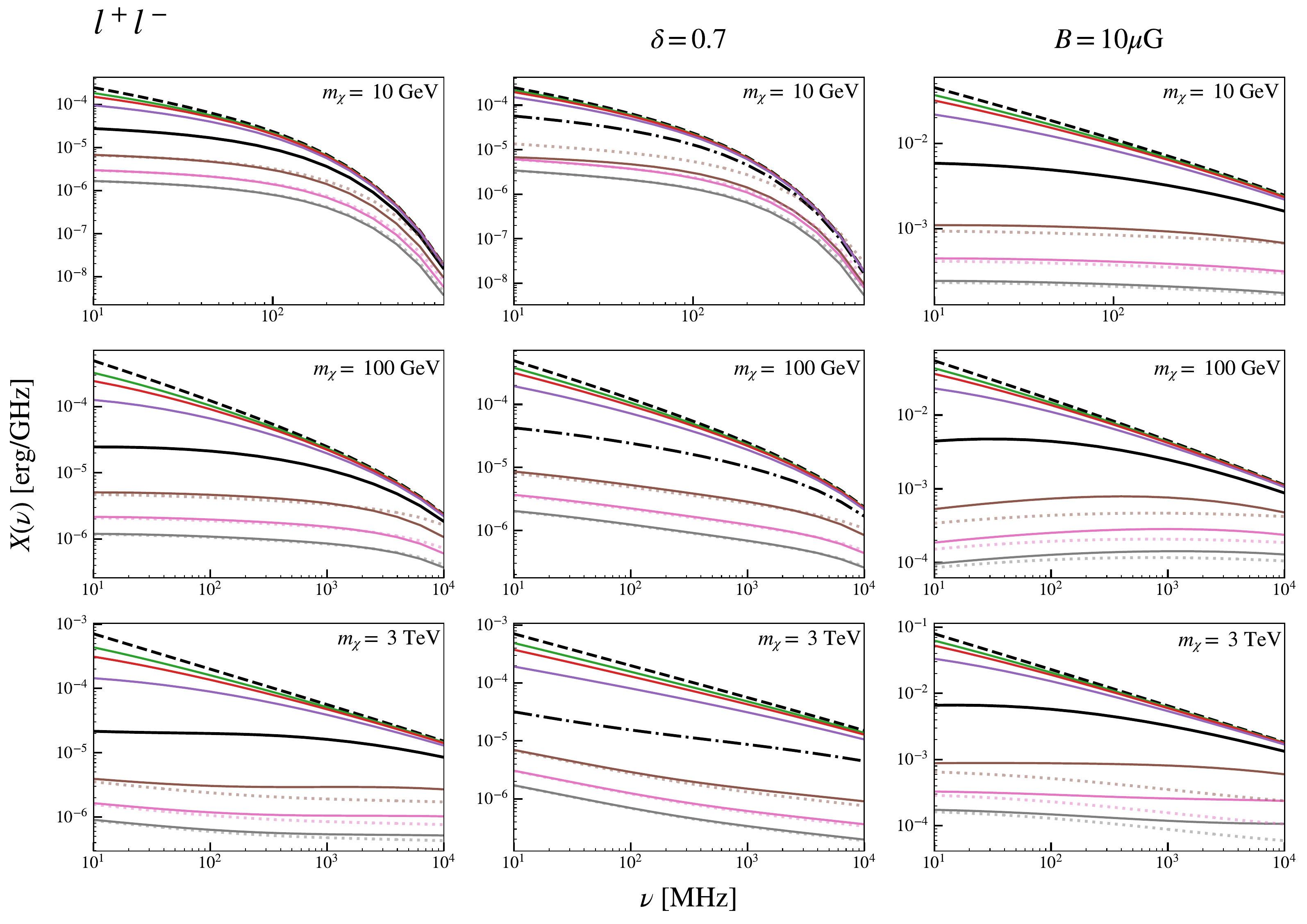}
\caption{\label{fig:ll} Spectral functions $\mx(\nu)$ ($l^+l^-$ final state) for several DM masses. Color and dashing as in fig. \ref{fig:eetree}.}
\end{figure}

\begin{figure}
\includegraphics[width=\linewidth]{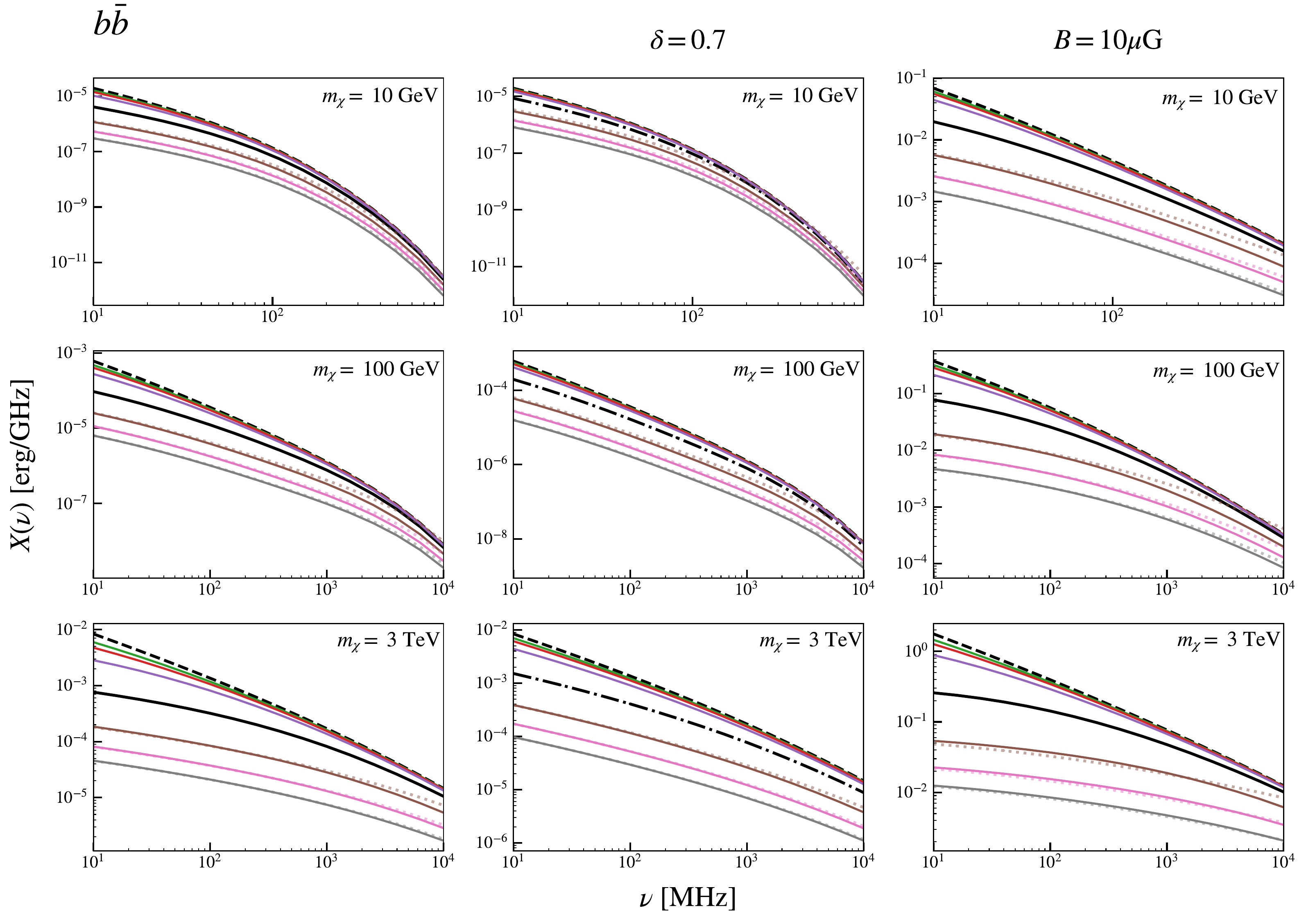}
\vspace{10pt}
\hrule
\vspace{20pt}
\includegraphics[width=\linewidth]{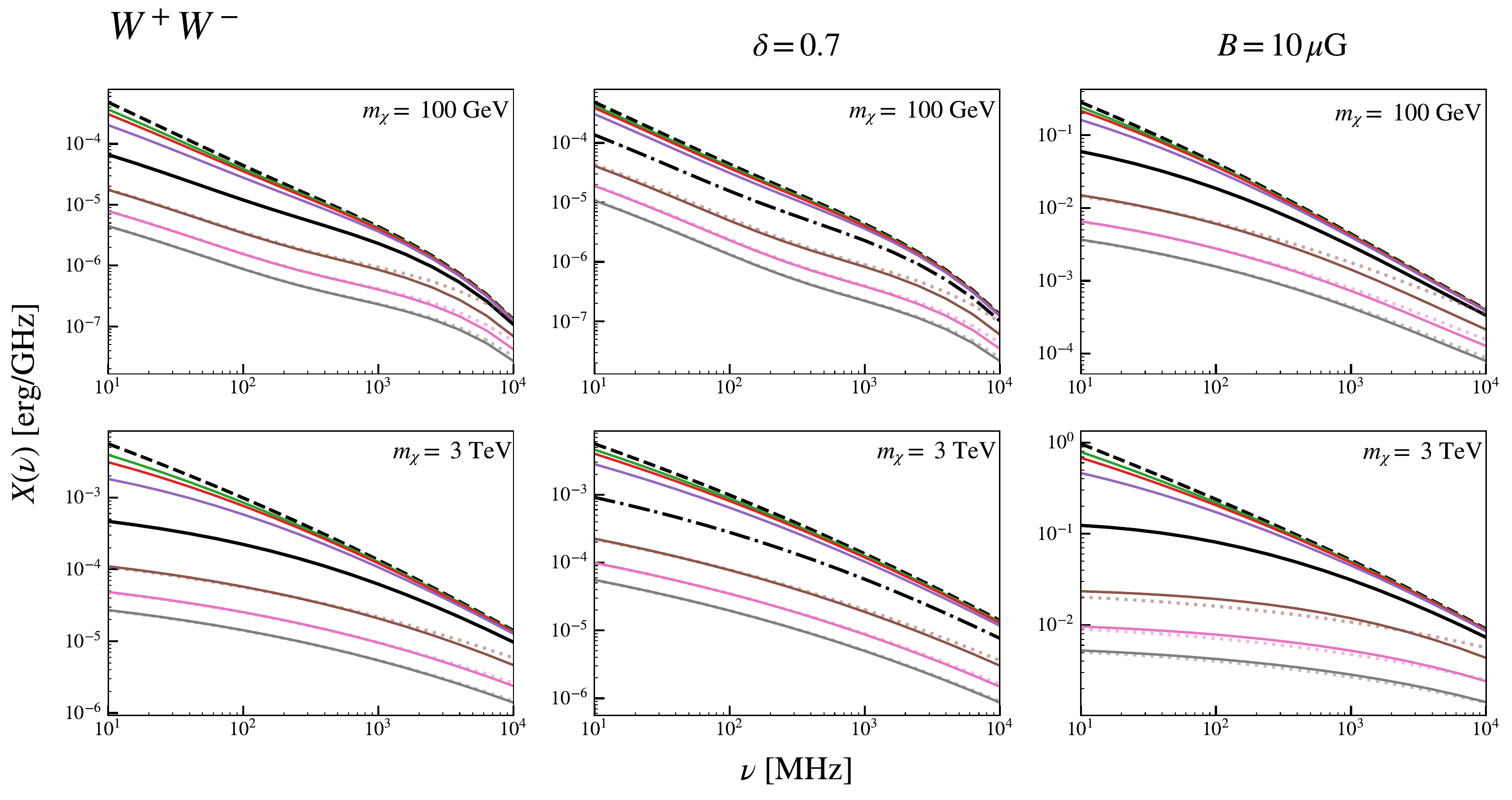}
\caption{\label{fig:WWbb} Same as fig. \ref{fig:ll} but for the $W^+W^-$ and $b\bar b$ final states.}
\end{figure}

In this appendix we show the spectral functions $\mx(\nu)$ defined in \eqref{eq:specb} for several realistic particle DM models.
Namely, in figs. \ref{fig:ll}-\ref{fig:WWbb} we consider the same parameter sets as in fig. \ref{fig:eetree} but more sophisticated electron yields.
Concretely, we assume in fig \ref{fig:ll} that the particle DM model is such that the annihilation/decay produces all charged leptons with the same probability.
In fig. \ref{fig:WWbb}, instead, $b$-quark or $W$-boson particle-antiparticle pair production is the dominant hard process in the annihilation/decay event.

\section{Further tests}
\label{app:bench}
In section \ref{sec:exact} we showed how the different approximations that we have put forward in this paper compare with those predictions that result from the full treatment.
Specifically, in fig. \ref{fig:exact} we performed such comparisons for three benchmark points represented as blue crosses in fig. \ref{fig:contours}.
These points gave us insights on the validity of the NDA (RA formula) and the RA-RC transition. 
However, the also interesting RA-RB transition and the validity of the LMA (RB formula) and the RDA (RC formula) was not covered there. 
In this appendix we consider additional benchmark points that precisely sit on those regions.
They are indicated by the orange circles.
Concretely, for leptophilic 50~GeV DM we consider a RA-RB transition point such that $\tau_0=4$~Gyr. 
For the same DM model we also consider a RB point such that $\tau_0=0.6$~Gyr.
Our third point, 2~TeV DM ($b\bar b$), is in the rapid-diffusion regime (C), $\tau_0=0.2$~Gyr. 

\begin{figure}
\includegraphics[width=\linewidth]{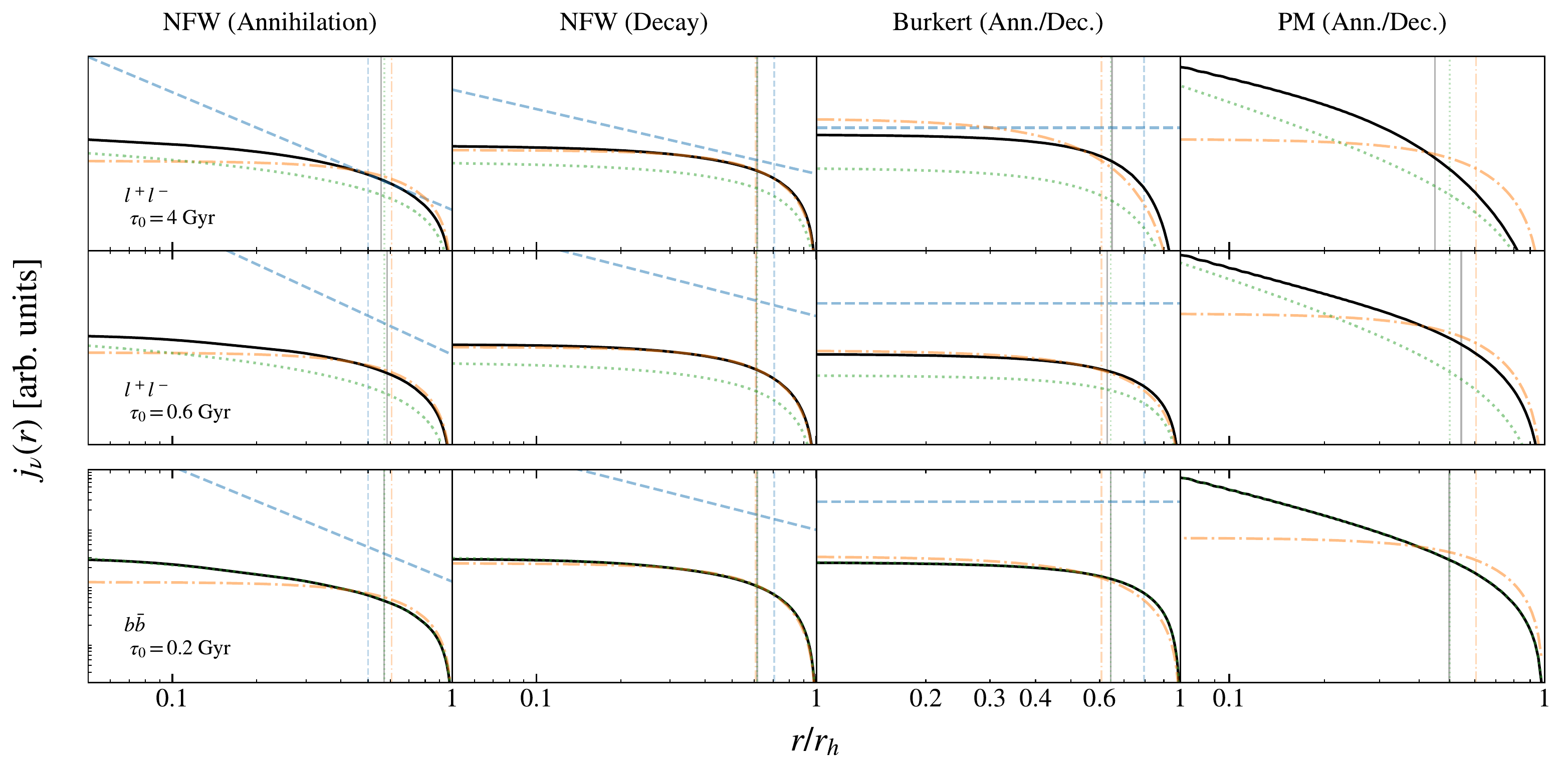}
\caption{\label{fig:benchapp} Emissivity profiles for the benchmark scenarios indicated with orange circles in fig \ref{fig:contours}. 
Concretely, a leptophilic model with $\mchi=50$~GeV (ann.) in the first and second rows. 
In the third row we consider a model with a DM particle of mass $\mchi=2$~TeV (4~TeV) such that it predominantly self-annihilates (decays) into $b\bar b$ pairs.
Color and line style as in fig. \ref{fig:exact}.}
\end{figure}

The results of these additional comparisons are shown in fig. \ref{fig:benchapp} (see also fig. \ref{fig:exact} for reference).
These agree with our expectations.
Namely, the black lines in the upper panels share common features with both the RA (dashed blue) and RB (dot-dashed orange) computations. 
With the exception of the point-mass DM profile, the latter approximate much better the exact results, as obvious from the figure (see also the vertical lines). 
Numerically, the RA flux densities are larger than the exact ones by $\sim$ 160\% (Burkert), 110\% (NFW/decay), 60\% (NFW/annihilation) and 12\% (point mass).

In the purely B and C regimes, the agreement between the approximations and complete computations is remarkable.
The panels in the middle are such that the differences between the flux-density computations are at most of $\mathcal O$(10\%) for point-mass profiles and below 1\% for NFW (decay).
Similarly, the lower panels (regime C) show differences of at most 1\% in all cases.

\section{Method-of-images formula}
\label{app:moi}
The solution of the transport equation \eqref{eq:diffloss} in terms of the Green's function \eqref{eq:fourier} is the main result of this work.  
Though, it would be interesting to contrast this picture with the strategy that has traditionally been employed so far, i.~e. integrating \eqref{eq:masterdiffloss} using the first few terms of the method-of-images (MoI) solution \eqref{eq:images}.
In this appendix we will perform comparisons between these two strategies.
In particular, we consider for simplicity those cases where the $r'$ integral in \eqref{eq:masterdiffloss} (using \eqref{eq:images} for the Green's function) can be analytically computed: point-mass, constant-density and NFW profiles.
In the latter case we assume that $r_s\to\infty$ limit and that DM decay is the process of interest.

\begin{figure}
\includegraphics[width=\linewidth]{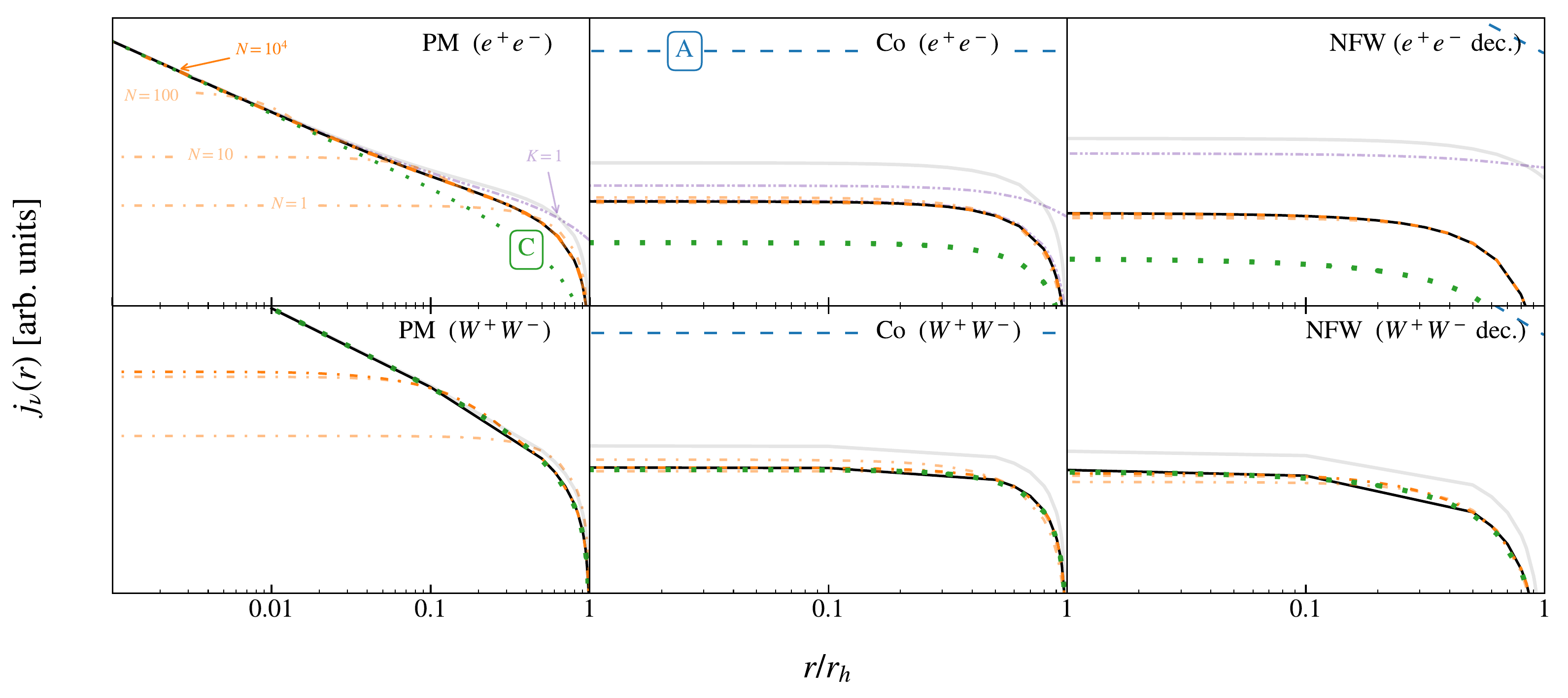}
\caption{\label{fig:images} Emissivity predictions ($\nu=150$~MHz and benchmark transport parameters) using \eqref{eq:masterdiffloss} with the MoI representation of the Green's function (solid black) and the Fourier one (dot-dashed orange) for $\mchi=100$~GeV (upper panels) and 6~TeV (lower panels) DM particles decaying respectively into $e^+e^-$ and $W^+W^-$ pairs. 
Dot-dot-dashed purple lines show contributions of the leading term in the MoI series.
Correspondingly, lighter dot-dashed orange lines represent the incomplete Fourier series (truncation order indicated in each curve). 
Dashed blue and dotted green curves are obtained using the A and C approximation respectively.
Light gray curves are instead obtained using the \old{} formula.
}
\end{figure}

The formulas are shown below and plotted in fig. \ref{fig:images} for the monochromatic electron injection ($\mchi=100$~GeV for DM decay) and a much heavier DM particle ($\mchi=6$~TeV) that decays primarily into a $W^+W^-$ pair.
The benchmark set of transport-model parameters and the frequency $\nu=150$~MHz is adopted in the computations. 
For the sake of generality we omitted the units on the (logarithmically scaled) emissivity axis. 
However, since some quantitative discussions are in order, we shall refer to some ratios between the several results included in the figure in our examinations below.

Specifically, the results displayed in fig. \ref{fig:images} demonstrate that indeed the formulas \eqref{eq:images} and \eqref{eq:fourier} are equivalent. 
However, the regimes of convergence are different. 
More concretely, we see that for the PM model, the MoI solution reproduces (at its leading order $m=0$) the small-$r$ behaviour of the emissivity while the Fourier representation requires infinite modes in order to reproduce the expected $r^{-1}$ singularity of the full solution. 
Instead, the \eqref{eq:fourier} (or \eqref{eq:emissexact}) emissivity features a core whose radius is inversely proportional to  $N$, the value at which the Fourier series is truncated in our computations.
The physical interpretation is that the $N$-th order Fourier expansion \eqref{eq:emissexact} can satisfactorily describe length scales of the system that are larger than $k_N^{-1}\sim\mathcal O(r_h/N)$, where $k_N$ is the wavenumber that is associated with the $N$-nth mode.
As discussed above, a similar phenomenon also happens in the case of DM annihilation in an NFW halo. 
In that case the singularity is much shallower $\ln(r)$.
Nevertheless, in practice this limitation of \eqref{eq:emissexact} is irrelevant because of the finiteness of the angular resolution of the telescopes.

On the other hand, the convergence of the MoI formula is worst at larger values of $r$ and for all $r$ in the constant DM density and NFW models. 
In particular, the boundary condition at $r=r_h$ is satisfied only approximately.
We also compared our results with those that are obtained by the \old{} Green's function
\begin{eqnarray}
\frac{r'}rG_{r_h}^{\rm old}(\Delta\lambda^2,r,r')&=&\frac{\Theta(\Delta\lambda^2)}{\sqrt{4\pi\Delta\lambda^2}}\lim_{L\to\infty}\sum_{l=-L}^L\left.\frac{r'}{r_l}(-1)^l\left(\e^{-\frac{(r'-r_l)^2}{4\Delta\lambda^2}}-\e^{-\frac{(r'+r_l)^2}{4\Delta\lambda^2}}\right)\right|_{r_l=(-1)^lr+2lr_h}\nonumber\\
\label{eq:oldmoi}
{}&=&\frac{\Theta(\Delta\lambda^2)}{\sqrt{4\pi\Delta\lambda^2}}\lim_{K\to\infty}\sum_{k=-K}^K\frac{r'}{r-2kr_h}(-1)^k\left(\e^{-\frac{(r-r'-2k r_h)^2}{4\Delta\lambda^2}}-\e^{-\frac{(r+r'-2kr_h)^2}{4\Delta\lambda^2}}\right)\ ,
\end{eqnarray}
where in the last step the terms have been regrouped in order to ease the comparison with \eqref{eq:images}. Considering that the $k=0$ term of this series agrees with \eqref{eq:images}, we find that the \old{} formula tends to deviate less from the correct result when cusped (PM and NFW/ann.) profiles are considered. 
However, differences are typically much more pronounced for shallower halo profiles.
Numerically, we find that the \old{} prediction is a factor $\sim$3 larger than the correct answer in the case of the constant-density model.
For the NFW model (decay), we find that the \old{} formula does not converge. 
The (gray) curves on the rightmost panels have been obtained by truncating the series \eqref{eq:oldmoi} at the value $L=10$.

\subsection{Point mass}
Assuming for concreteness DM annihilation, it is straightforward to obtain the MoI solution for a point-mass DM halo function. Namely,
\begin{eqnarray}
n_e^{\rm PM}(r,E)&=&\frac{\sv K_{\rm pm}}{2\mchi^2b(E)}\sum_{k=-\infty}^{\infty}\frac{r-2kr_h}{(4\pi\Delta\lambda^2(E))^{3/2}r}\e^{-\frac{(r-2kr_h)^2}{4\Delta\lambda^2(E)}}\nonumber\\
{}&\simeq&\frac{\sv K_{\rm pm}\e^{-\frac{r^2}{4\Delta\lambda^2(E)}}}{2\mchi^2b(E)(4\pi\Delta\lambda^2(E))^{3/2}}\left(1+2\e^{-\frac{r_h^2}{\Delta\lambda^2(E)}}\cosh\left(\frac{r_h r}{\Delta\lambda^2(E)}\right)
\right.\nonumber\\
{}&{}&\left.
-\frac{4r_h\e^{-\frac{r_h^2}{\Delta\lambda^2(E)}}}{r}\sinh\left(\frac{r_h r}{\Delta\lambda^2(E)}\right)+\ldots\right)\ .
\end{eqnarray}

For DM decay, the formula is analogue: $\sv K_{\rm pm}/(2\mchi^2)$ should be substituted by $M_h/(\mchi\tau)$.

\subsection{Constant-density profile}
In the case of a cored profile and DM annihilation, the MoI solution is given by 
\begin{eqnarray}
n_e^{\rm C}(r,E)&=&\frac{\sv\rho_h^2}{2\mchi^2 b(E)}\left\{1+\frac{r_h}{r}\lim_{K\to\infty}\sum_{k=-K}^{K+1}{\rm erf}\left(\frac{(2k-1)r_h-r}{2\Delta\lambda(E)}\right)\right\}\nonumber\\
&=&\frac{\sv\rho_h^2}{2\mchi^2 b(E)}\left[1+\frac{r_h}{r}\left({\rm erf}\left(\frac{r_h-r}{2\Delta\lambda(E)}\right)-{\rm erf}\left(\frac{r_h+r}{2\Delta\lambda(E)}\right)+\ldots\right)\right]\ .
\end{eqnarray}
As before, the result is similar had DM decay been our process of interest instead of annihilation: $\sv \rho_h^2/(2\mchi^2)$ should be substituted by $\rho_h/(\mchi\tau)$.

\subsection{NFW profile}
The solution \eqref{eq:masterdiffloss} for decaying DM in an NFW halo and in the $r_s\gg r_h$ limit is given by
\begin{equation}
\label{eq:moinfw}
n_e^{\rm NFW}(r,E)=\frac{4\rho_sr_s}{b(E)\mchi\tau}\frac1{r}\lim_{K\to\infty}\sum_{k=-K}^{K}(-1)^k{\rm erf}\left(\frac{kr_h+r}{2\Delta\lambda(E)}\right)\ .
\end{equation}
In contrast with the former two cases, there is no simple analytical solution if DM annihilation is the process of interest.

\section{RMS noise and signal upper bounds}
\label{app:bounds}
In this appendix we derive a very simple formula that allows the user to estimate upper bounds on e.~g. the annihilation rate $\langle\sigma v\rangle$ in some WIMP model, that can be placed by the non-observation of a given dwarf galaxy.
Specifically, we would like to prove that by using the non-observation root-mean-square (RMS) noise level $\Phi_{\rm rms}$, the telescope beam size $\Omega_B $ and the total area of the image $\Omega_T$, 2$\sigma$ upper bounds on a given signal prediction $\Phi_S$ can be obtained. 
These are given by
\begin{equation}
\Phi_S\lesssim\sqrt{2.71\times\frac{\Omega_B}{(1-f)f\Omega_T}}\Phi_{\rm rms}\simeq1.64\frac{\Phi_{\rm rms}}{\sqrt{N_{\rm beams}}}\ ,
\label{eq:appbounds}
\end{equation}
where $f\Omega_T$ is the angular extension of the signal and, thus, $N_{\rm beams}\equiv f\Delta\Omega_T/\Delta\Omega_B$ is the effective number of beams that are necessary to enclose the signal. 
The last equality is of course valid provided $f\ll1$.

In order to prove \eqref{eq:bounds}, we first discretize the image and the data.
Specifically, we assume that for every pixel $i=1,\ldots,N$ the telescope measures a ``number'' $X_i$ of events after some given time. 
Each element in the dataset $\{X_i\}_{i=1}^N$ is also assumed to follow a Poisson process with a Poisson parameter $\lambda$ common for all $i$'s.
In the high-statistics limit (assumed throughout) the root mean square (RMS) deviation of the whole data $\sigma_{\rm rms}$ is given by $\sqrt\lambda$ and also $\lambda=(1/N)\sum_{i=1}^NX_i$. 

Next, we pose the following problem: given a top-hat signal model $\mx_i=S$ if $i\leq f N$ and zero otherwise, what are the bounds on the parameter $S$ that can be obtained by fitting this signal (plus background) to the noisy data $\{X_i\}_{i=1}^N$? 
The background model $B\approx\lambda$ is assumed to be the same for all pixels.

In order to address the above question, we perform a likelihood ratio test on the model.
Let $C=S+B$ so that the likelihood function is given by
\begin{equation}
\mathcal L(S,B)=\prod_{i=1}^{fN}\frac{C^{X_i}\e^{-C}}{X_i!}\times\prod_{j=fN+1}^{N}\frac{B^{X_j}\e^{-B}}{X_j!}\ .
\end{equation}

Since the subsets $\{X_i\}_{i=1}^{fN}$ and $\{X_j\}_{j=fN+1}^N$ are statistically independent, the maximum likelihood is reached when $C=B=\lambda$ ($\hat S=0$). 
Thus,
\begin{equation}
\frac{\mathcal L(S,B)}{\mathcal L_{\rm max}}=\left(\frac{B+S}{\lambda}\right)^{fN\lambda}\e^{-fN(B+S-\lambda)}\left(\frac{B}{\lambda}\right)^{(1-f)N\lambda}\e^{-(1-f)N(B-\lambda)}\ ,
\label{eq:likeratio}
\end{equation}
and the test-statistics function is given by
\begin{equation}
{\rm TS}=-2\log\frac{\mathcal L(S,B^*)}{\mathcal L_{\rm max}}=2N\left\{B^*+fS-\lambda-f\lambda\log\left(\frac{B^*+S}{\lambda}\right)-(1-f)\lambda\log\left(\frac{B^*}{\lambda}\right)\right\}\ ,
\label{eq:ts}
\end{equation}
where $B^*$ is the value that, for fixed $S$, minimizes \eqref{eq:likeratio}:
\begin{equation}
B^*=\lambda-fS+f(1-f)\frac{S^2}{\lambda}+\mathcal O(S^3/\lambda^2)\ .
\end{equation}

After a rather lengthly expansion in $S$ of the terms of \eqref{eq:ts} where all terms of $\mathcal O(S^0)$ and $\mathcal O(S^1)$ cancel with each other, we obtain
\begin{equation}
{\rm TS}=(1-f)fN\frac{S^2}{\lambda}+\mathcal O(S^3/\lambda^2)\ .
\end{equation} 
The $2\,\sigma$ upper bound is thus obtained by substituting TS~$=2.7055\ldots$ in the above equation. 
This value for TS solves the equation $1-\Gamma(1/2,{\rm TS}/2)/\sqrt\pi=90\%$ and it corresponds to the one-sided exclusion of the model at a 95\% confidence level.
Therefore,
\begin{equation}
S_{2\sigma}=\sqrt{2.7055\ldots}\frac{\sigma_{\rm rms}}{\sqrt{f(1-f)N}} \ ,
\end{equation}
QED.

%
\end{document}